\documentclass{aa}  

\usepackage{amsmath}
\usepackage{bm,ulem}
\usepackage{graphicx}
\usepackage{siunitx}

\usepackage{color}

\usepackage{txfonts}
\usepackage[breaklinks=true]{hyperref}
\hypersetup{colorlinks=true,linkcolor=blue,citecolor=blue,urlcolor=blue}

\newcommand\gaia{\textit{Gaia}}
\newcommand\gdr[1]{\gaia~DR#1}         % used in release documentation
\newcommand\egdr[1]{\gaia~EDR#1}

\providecommand{\dt}[1]{{\tt #1}} % for field names in the data model

\begin{document} 

   \title{\gaia\ Early Data Release 3}

   \subtitle{Building the \gaia\ DR3 source list --- Cross-match of \gaia\ observations}

% Reference person for this section: FTC

% 14 names as of November 27

\author{F.~Torra\inst{\ref{inst:dapcom}}\fnmsep\thanks{Corresponding author: F.~Torra\newline e-mail: \href{mailto:ftorra@fqa.ub.edu}{\tt ftorra@fqa.ub.edu}} % as main writer and actual leader of IDU-XM
\and J.~Casta\~neda\inst{\ref{inst:dapcom}} % as writer and actual manager of IDU
\and C.~Fabricius\inst{\ref{inst:ieec}} % as co-ordinator, important contributor to many aspects
\and L.~Lindegren\inst{\ref{inst:lund}} % for major contributions in the XM clustering algorithm
\and M.~Clotet\inst{\ref{inst:ieec}} % as previous XM leader
\and J.J.~Gonz\'alez-Vidal\inst{\ref{inst:dapcom}}  % for developments in the XM resolver algorithms
\and S.~Bartolom\'e\inst{\ref{inst:ieec}} % for operational tasks of XM
\and U.~Bastian\inst{\ref{inst:ari}} % as CU3 leader
\and M.~Bernet\inst{\ref{inst:ieec}} % for validation of XM
\and M.~Biermann\inst{\ref{inst:ari}} % as CU3 leader
%\and T.~Br\"usemeister\inst{\ref{inst:ari}} % as contributor to the HPM catalogue for DR2
\and N.~Garralda\inst{\ref{inst:ieec}}  % for developments in DC
\and A.~G\'urpide\inst{\ref{inst:ieec}} % for developments in the XM algorithms
\and U. Lammers\inst{\ref{inst:esac}} % as CU3-T and SOC manager
\and J.~Portell\inst{\ref{inst:ieec}} % as contributor to many aspects
%\and R.~Smart\inst{\ref{inst:oato}} % as contributor to the HPM catalogue for DR2
%\and A.~Spagna\inst{\ref{inst:oato}} % as XM contributor and former leader
\and J.~Torra\inst{\ref{inst:ieec}} % as manager of IDT/IDU, responsible for the Spanish contribution, and as CU3 deputy
}

\institute{
DAPCOM for Institut de Ci\`encies del Cosmos,
Universitat de Barcelona (IEEC-UB), Mart\'i i Franqu\`es 1, E-08028 Barcelona, Spain\label{inst:dapcom}
\and
Institut de Ci\`encies del Cosmos,
Universitat de Barcelona (IEEC-UB), Mart\'i i Franqu\`es 1, E-08028 Barcelona, 
Spain \label{inst:ieec}
\and
Lund Observatory, Department of Astronomy and Theoretical Physics, Lund University, Box 43, 22100, Lund, Sweden
\label{inst:lund}
\and
Astronomisches Rechen-Institut, Zentrum f\"ur Astronomie der Universit\"at Heidelberg, M\"onchhofstra{\ss}e 14, D-69120 Heidelberg,
Germany
%\ \email{bastian@ari.uni-heidelberg.de}
\label{inst:ari}
\and
ESA, European Space Astronomy Centre, Camino Bajo del Castillo s/n, 28691 Villanueva de la Ca{\~n}ada, Spain\label{inst:esac}
%\and
%Istituto Nazionale di Astrofisica, Osservatorio Astrofisico di Torino, Via Osservatorio 20, Pino Torinese, Torino, 10025, 
%Italy
%\ \email{busonero@oato.inaf.it;riva@oato.inaf.it;\\lattanzi@oato.inaf.it;drimmel@oato.inaf.it}
\label{inst:oato}
}

   \date{ }

% \abstract{}{}{}{}{} 
% 5 {} token are mandatory
 
  \abstract
   { The early \gaia\ Data Release 3 (\egdr{3}) contains results derived from 78 billion 
individual field-of-view transits of 2.5~billion sources collected by the European Space Agency's \gaia\ mission during 
its first 34 months of continuous scanning of the sky.}
   {We describe the input data, which have the form of onboard detections, and the 
modeling and processing that is involved in cross-matching these detections to sources. For the cross-match, we formed clusters of detections that were all linked to the same physical light source on the sky.}
   {As a first step, onboard detections that were deemed spurious were discarded.
The remaining detections were then preliminarily associated with one or more sources
in the existing source list
in an observation-to-source match. 
All candidate matches that directly or 
indirectly were associated with the same source form a match candidate group. The detections from the same group were then subject to a cluster analysis. Each 
cluster was assigned a
source identifier that normally was the same as the identifiers from \gdr{2}. 
Because the number of individual detections is very high, we also describe the efficient organising of 
the processing.}
   {We present results and statistics for the final cross-match with particular 
emphasis on the more complicated cases that are relevant for the users of the \gaia\ 
catalogue. We describe the improvements over the earlier \gaia\ data 
releases, in particular for stars of high proper motion, for the brightest 
sources, for variable sources, and for close source pairs.}
   {}


   \maketitle
%
%-------------------------------------------------------------------

\section{Introduction}\label{S:introduction}

% Reference person for this section: CF / FTC

% This section should introduce the problem we want to tackle at a high level and the requirements for Gaia XM. \citet{Fabricius2016}

Since 2014, the \gaia\ mission of the European Space Agency has been carrying
out an all-sky astrometric and photometric survey \citep{Prusti2016}. The results of the first 34
months of mission are published in two stages.  The first is the early data
release, \egdr{3}\footnote{\url{https://www.cosmos.esa.int/web/gaia/earlydr3}} \citep{EDR3-DPACP-130}. It contains astrometry and photometry
for more than 1800 million sources. It is followed by the full \gdr{3},
which will add many more data aspects, such as astrophysical parameters and updated
radial velocities, to the same source list.

\gaia\ is scanning the sky in great circles that on average cover any point
on the sky more than a dozen times per year.  The \gaia\ instrument includes two
telescopes, separated by 106.5\degr, with a common focal plane. A sufficiently
point-like source, which is transiting the field of view (FoV) of the telescopes, is
detected by a sky mapper (SM) and subsequently observed in the astrometric
(AF) and photometric (BP and RP) fields, and sometimes also by the spectroscopic instrument (RVS). 

\begin{figure}[h]
\centering
\includegraphics[width=0.9\columnwidth]{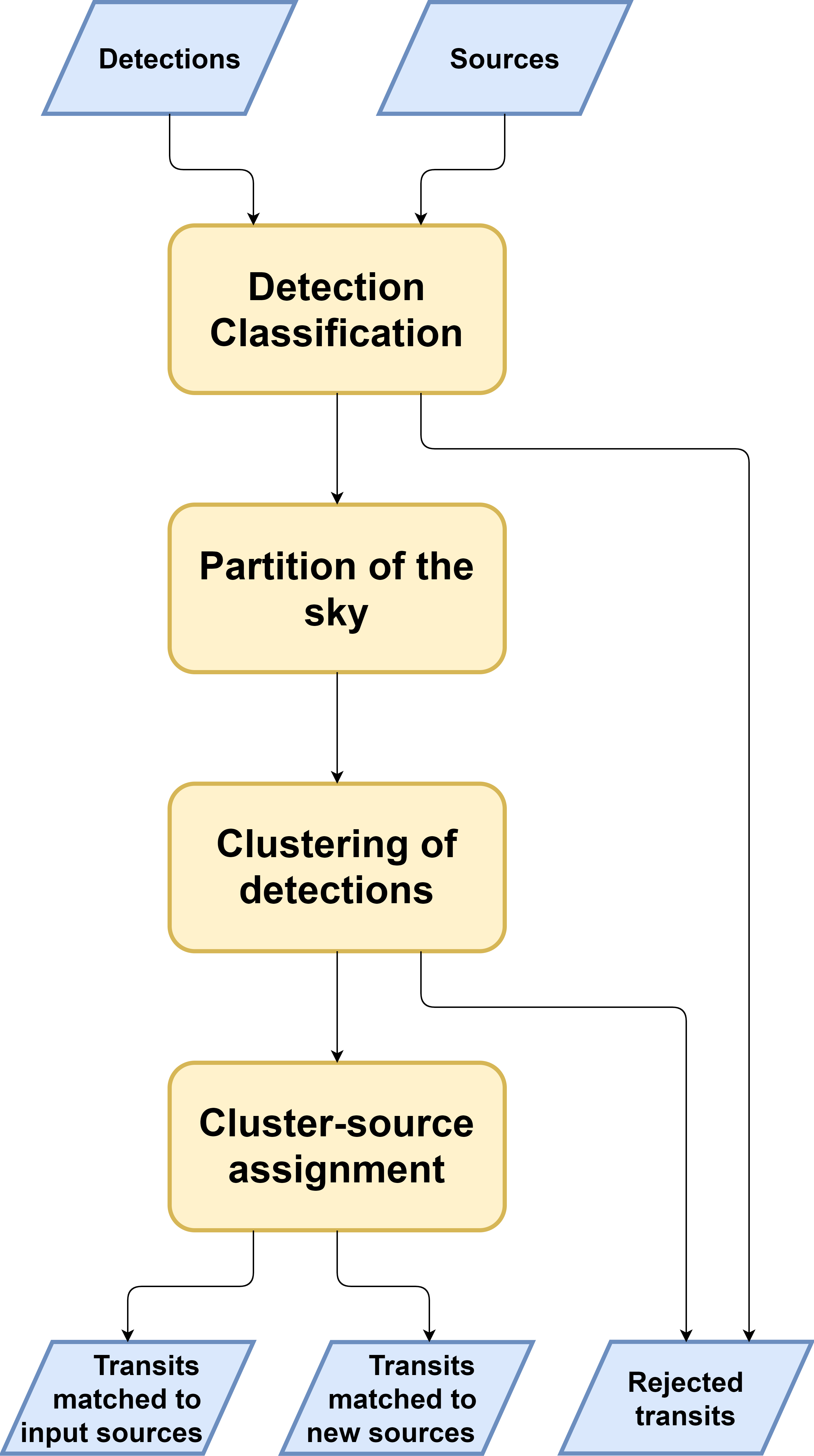} 
\caption{Flow diagram of the XM process. The XM uses the onboard detections and catalogue sources with the latest
calibrations and parameters from the previous \gaia~DR. The first stage is the detection classifier
to distinguish real from spurious detections. It is mainly based on detection information, but uses the position of bright sources to classify detections around them. Then the sky is partitioned into groups of detections and sources.
These groups are processed without using any source information to determine clusters of detections
that belong to the same source. Finally, each cluster is assigned to an input source, or a new source
is created. XM provides a list of matched transits to already known and new sources to the subsequent pipelines,
together with a list of rejected transits.
\label{fig:fd}}
\end{figure}

The aim of the cross-match (XM) process that we describe here is to identify all onboard
detections belonging to the same physical light source on the sky and assign a unique identifier to each
such cluster of detections and thus to the source. The logical structure of the XM is summarised in Fig.~\ref{fig:fd}. For the first two \gaia\ data releases, the XM
process was described by \citet{Fabricius2016}, with updates for \gdr{2} by
\citet{Lindegren2018}.  These two papers had a different focus and only
included a top-level description of the XM process. The present
paper is motivated by the need to provide a more detailed exposition  
and to describe the improvements for \egdr{3 as well}. We would like to point out that the source list will be the same for \egdr{3} and \gdr{3,} and thus the XM process applies to both releases.

The XM process described here is the full-scale XM creating the source list for \egdr{3}. This is a complex process that involves
78 billion detections of 2.5 billion sources and covers nearly three
years of observations.  A simpler XM method is used for the daily processing
of newly arrived data. It is described in \citet{Fabricius2016} and is not
included here.  We do not discuss the identification of Solar System
objects either, which is handled by a separate pipeline \citep{2018A&A...616A..13G}.
We therefore treat them just like any other detections and ignore their peculiar nature. Typically, the detection of a Solar System
object therefore internally gives rise to a cluster with a single
detection and a source identifier of its own, but these single detections are
discarded for the data release and are picked up independently by the dedicated
pipeline. 

The actual data release contains fewer than 2.5 billion sources because some
sources have not yet been observed a sufficient number of times to allow for a proper astrometric and photometric solution. This mostly
occurs close to and beyond the intended detection limit of $G\sim 20.7$~mag.

The XM table that links transit identifiers with source identifiers is
the same for astrometry, photometry, and spectroscopy. A transit identifier unequivocally labels a field-of-view (FoV) transit, which contains a set of one SM observation, up to nine AF observations, and usually one BP and one RP observation each. A transit
identifier is based on the result of an onboard SM detection (in the following called a `detection'). Such a
detection is expressed in terms of a `reference acquisition pixel' (giving a rough estimate of the location of the first AF window on the sky)
and of a $G$-band magnitude, and it controls the onboard data acquisition.  We
decided to use this reference pixel from the onboard detection to associate
each detection with a sky position because this simple and robust strategy will work regardless of possible complications of an astrometric observation. In general, the astrometric observations are more accurate
in the scanning direction, while only the reference pixel gives (moderately) accurate
information for the across-scan (AC) direction.  The drawback of using the reference
pixel is that the relatively large position errors for the brightest sources
($G <8$~mag) go uncorrected into the transit identifier, and close source pairs remain unresolved at
XM level.

The clusters of detections constructed in the match process are assigned unique
source identifiers. These source identifiers coincide with those
used in the previous data releases for the common set of sources. Because the input set of detections grows between processing cycles, inconsistencies between
sources cannot always be avoided. It may happen that at a position
where we previously had two `sources', we now only have one, or the other way
around. We refer to these cases as `merge' and `split', respectively. They lead to the
assignment of new source identifiers. As intermediate cases, detections may
migrate between neighbouring clusters, but this does not lead to a revision of the
source identifier.  

In the following sections we describe the dataset in more detail
(Sect.~\ref{S:data_used}), discuss the way spurious detections are identified
and rejected (Sect.~\ref{S:dc}), the way the XM task is broken down
to process-independent patches of the sky (Sect.~\ref{S:mcg}), present the mathematical model
for forming clusters of detections (Sect.~\ref{S:clustering}), the changes in the
source list between processing cycles (Sect.~\ref{S:src_assignation}), the
validation of the results (Sect.~\ref{S:validation}), and finally give an outlook on the
developments for the processing for future data releases
(Sect.~\ref{S:conclusions}).

\section{Input data}\label{S:data_used}

% Reference person for this section: JC / FTC

\begin{figure}[h]
\centering
\includegraphics[width=0.85\columnwidth]{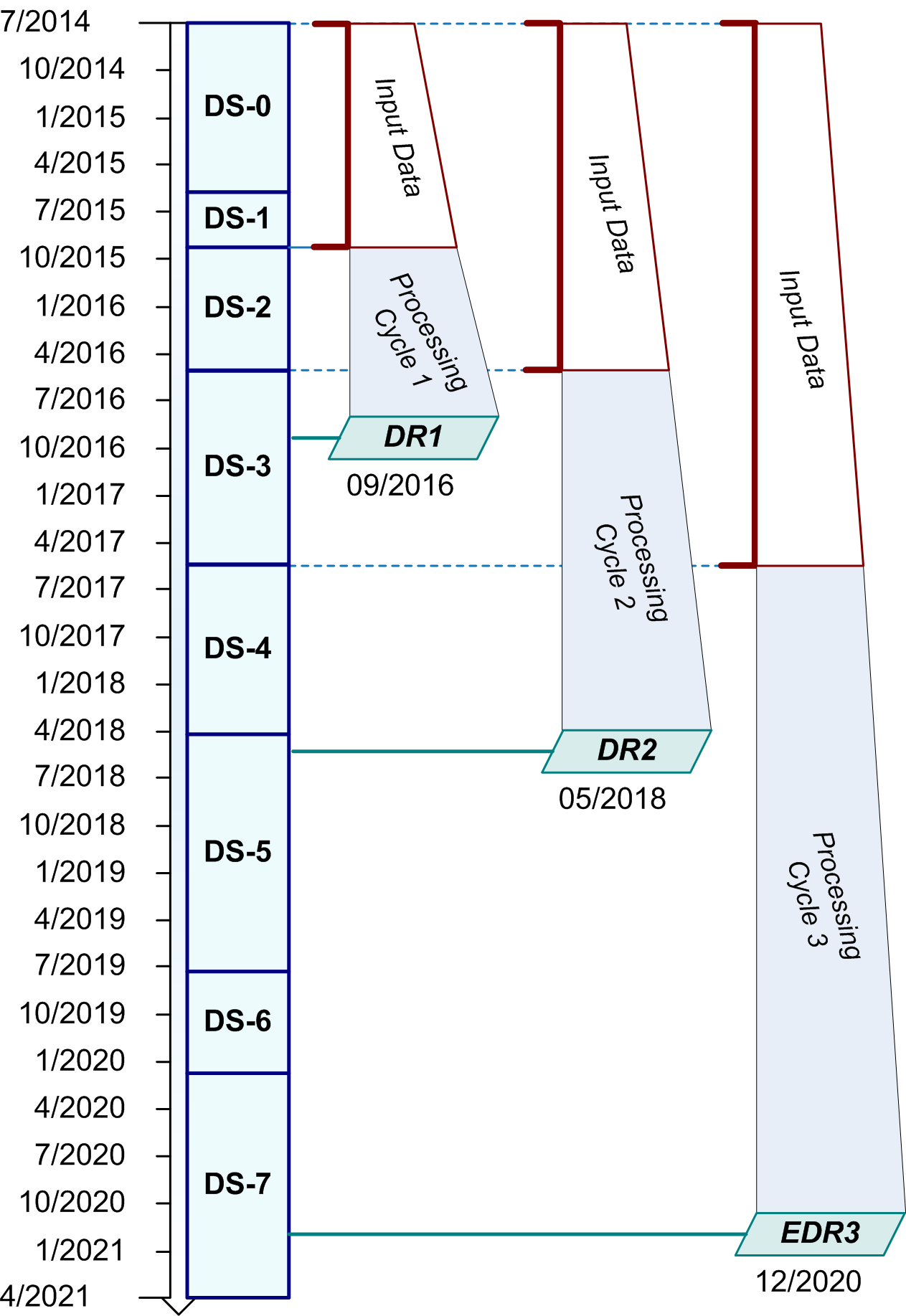} 
\caption{Schematic diagram of the data segments since the start of nominal operations in July 2014 up to the \egdr{3} publication.
New data are received on a daily basis from the spacecraft and continuously enter the \gaia\ data-processing pipeline.
This data stream is partitioned into data segments to identify and coordinate the data entering each processing cycle from which
\gaia\ DRs are produced. In each processing cycle, all accumulated data segments are processed again, superseding any preceding solution. DS-0 covers 313~days and 22.2~billion detections, DS-1 covers 105~days and 8~billion detections, DS-2 covers 250~days and 22.1~billion detections, and DS-3, which is the new data segment for \egdr3, covers 370~days and 25.7~billion detections.
\label{fig:dataSegmentCycleAndGdr}}
\end{figure}

The main input data for the XM process are the onboard detections by the two \gaia\ sky mappers
(one for each telescope, and consisting of 14 CCDs in total).
This data stream is divided into data segments, lasting from 3 to 12 months, which are used to identify
the data entering in each processing cycle and thus each \gaia\ data release, as shown in Fig.~\ref{fig:dataSegmentCycleAndGdr}.
These data segments are processed several times during the mission at different data processing centres,
each with specific tasks assigned. For \egdr{3}, the data cover 1038 days (corresponding to the first four data segments),
from 25 July 2014 until 28 May 2017, with short breaks due to maintenance activities and spacecraft events. The density of detections is reduced after DS-2 because the parameters of the onboard detection algorithm change.

From these onboard detections, the sky coordinates of the observed objects are derived using the latest geometrical calibration of the focal plane
and refinements of the spacecraft attitude. The details of the sky-coordinate determination is described
in \citet[][Sect.~6.4]{Fabricius2016} and is not repeated here, but we describe in Sect.~\ref{SS:input_obs}
the parameters and statistical properties of the detections and the implications for their treatment in the cluster analysis.

The XM process also benefits from an input source catalogue (hereinafter referred to as the working catalogue) by keeping  the identifiers created in previous processing cycles as far as possible.
In Sect.~\ref{SS:src_catalogue} we identify the small set of source parameters of interest for XM, and we detail the origin of
and the update procedure applied on their values throughout the processing cycles. Finally, we  distinguish
between the contents of the source list we used for the cyclic processing and the source list that was published as part of the \gaia\ DRs.

\subsection{Detections}\label{SS:input_obs}

For each individual onboard detection, that is, for each detected FoV transit, the XM process derives the sky coordinates using the reference
acquisition pixel coordinates in AF1 (the first strip of astrometric CCDs). These reference pixel coordinates are
computed on board by propagating the transit time and AC column detected in SM to the expected position in AF1.
Based on this information, each transit reconstructed on ground is given a unique identifier or transitId.
This transitId is basically a numeric field coding the reference pixel coordinates in AF1, the telescope, and the CCD row
in which the object was observed.

The precision of the reference pixel coordinates is limited by the SM pixel resolution and by the precision
of the onboard image parameter determination.
As described in \citet{Prusti2016}, each pixel is 10~$\mu$m $\times$ 30~$\mu$m
(corresponding to 58.9~mas $\times$ 176.8~mas on the sky), whereas the SM has a reduced spatial
resolution with an on-chip binning of 2 pixels along-scan (AL) by 2 pixels AC per sample. The positional error used in XM therefore is a few tenths of an arcsec, depending on the scan direction and magnitude.
The decision to use the reference AF1 acquisition pixel was made because it is simple and robust: it already
provides sufficient accuracy for the XM purposes.
Together with the sky coordinates, the XM determines the scanning angle of the detection (i.e. the angle of
the great circle scanned by the spacecraft with respect to north at the position and time of the observation). This angle is 0\degr\
when scanning towards north, and 90\degr\ when scanning towards east.

However, there are several caveats regarding the reliability of the onboard \gaia\ detections.
Because of the limited resources available on board \gaia, the onboard confirmation algorithms applied for each object detected in the SM is different
depending on the estimated brightness and the acquisition mode \citep[][Sect.~2]{Fabricius2016}.
This confirmation process is even skipped in some cases, generating many unwanted spurious detections, mainly around bright sources
and for cosmic-ray events. These must be identified and discarded in the first step in the XM chain, as described in Sect.~\ref{S:dc}.

In addition to the sky coordinates, the XM also requires information about the detection brightness.
This information is mainly used to resolve the ambiguity in close-object  scenarios and crowded regions, but it is also
helpful for providing an initial estimate of the magnitude of new objects that are added to the catalogue for downstream processes.
The detection brightness used in the XM is the brightness measured onboard in the SM and has an estimated error of a
few tenths of a magnitude. This is also sufficiently accurate for our purposes.

\subsection{Source catalogue}\label{SS:src_catalogue}

The first XM run (on which \gdr{1} was based) started from the initial \gaia\ source list (IGSL) with 1200 million entries
compiled from the best optical astrometric and photometric catalogues of celestial objects available before the launch of \gaia\ \citep[][Sect.~6.1]{2014A&A...570A..87S,Fabricius2016}.
Each entry in this catalogue was given a unique identifier or sourceId. This sourceId is a numeric field to facilitate
identifying and spatially arranging it. This numeric field codes a spatial HEALPix index \citep{Gorski} and a running number.
The reference system for the \gaia\ source catalogue is the barycentric celestial reference system (BCRS/ICRS) \citep{Mignard2018}.

The working catalogue has been updated three times, one for each completed processing cycle and the corresponding \gaia~DR.
In each cycle, the number of entries has changed (either adding new entries or removing superseded entries), and
the source parameters have been replaced using the new astrometric and photometric solutions.
Two major updates must be highlighted regarding the completed two processing cycles preceding \egdr{3}.
The first update, in the very first cycle, when the number of entries was doubled compared to the initial source list, reached more than 2500~million entries.
The second update included the change of a significant fraction of the source identifiers at the bright end ($G<12$) \citep{Arenou2018}
and the manual recovery of thousands of stars with high proper motion (HPM) that were found to be missing in \gdr{1}.
There is a twofold explanation for the missing HPM stars: the initial source list did not contain sufficiently good
proper motion data for these sources, and secondly, the XM algorithm did not yet include a full-fledged
proper motion treatment and created several different (new) sources from subsequent transits of those stars.
After \gdr{1}, about 3000 new sources extracted from the revised Luyten half-second (LHS) catalogue \citep{Bakos2002} were
added to the working source catalogue in anticipation of the new XM run for \gdr{2}.

With this extension prepared by the Heidelberg ARI team and the algorithm improvements, the XM was able to match and
thus recover 2257 of these stars with a proper motion larger than 200~mas~yr${^{-1}}$.
The remaining unmatched HPM entries were not compatible with the available \gaia~detections, and they were discarded accordingly.

In addition to the update on the source identifiers, the source parameters may also show major updates depending
on the new detections entering the cycle and the reassignment of the detections done by the XM solution.
All these updates may complicate the source tracking in different \gaia~DRs, as explained in Sect.~\ref{S:src_assignation}.

The input working source catalogue for the \egdr{3} contains 2583~million sources, of which 1693~million sources were
published in \gdr{2} \citep{Brown2018},
Although a significant number of the non-published sources could be spurious, many of them are real sources with too
few matched detections to meet the publication quality thresholds.
When we discard these sources in the XM, we would create the same sources again until there finally are enough transits to allow a proper source creation, or unless the
detections that triggered their creation in the previous cycle vanish, which means that they become classified as spurious detections as described in Sect.~\ref{S:dc}.
Consequently, these sources were kept. This additionally helps to stabilise the source list between cycles.

We can distinguish several types of sources in the working catalogue according to their status and origin. Most of the sources have been updated by the global astrometric and photometric pipelines in the preceding cycle, but some of the input sources have not been updated by these two pipelines and contain obsolete solutions. Moreover, other input sources have been created in previous XM runs but never updated by any other pipeline across the processing cycles. Finally, there are surviving entries from the initial source catalogue with at least one matched detection but whose source parameters are external to the \gaia\ processing and have never been updated.

For each type of source, the XM applies a tailored treatment based on the available parameters, the producer, and its last update.
Consequently, we use the full set of parameters (including proper motion and parallax) for sources with a global astrometric solution
from the preceding cycle, whereas only the sky coordinates are considered for the other sources.
On the other hand, no distinction is made for the source magnitude because the threshold applied to this parameter during the source assignment
is more relaxed (see Sect. \ref{S:src_assignation}).

\section{Detection classification}\label{S:dc}

The \gaia\ onboard detection software autonomously distinguishes point-like source images from false detections
using  parametrised criteria of the image shape. However, when a high detection probability
for sources at the \gaia\  limiting magnitude of 20.7 mag is reached, it implies that the number of surviving spurious detections increases significantly.
A study of the onboard detection capability is provided by \citet{deBruijne2015}.

The spurious detections increase the complexity of the XM and may lead to weird clustering
solutions and incorrect source assignments.
When the spurious detections
are not clustered with other detections, they may trigger the creation of new sources in the XM solution, which will pollute the source catalogue.

\begin{figure}[h]
\centering
\includegraphics[width=1.00\columnwidth]{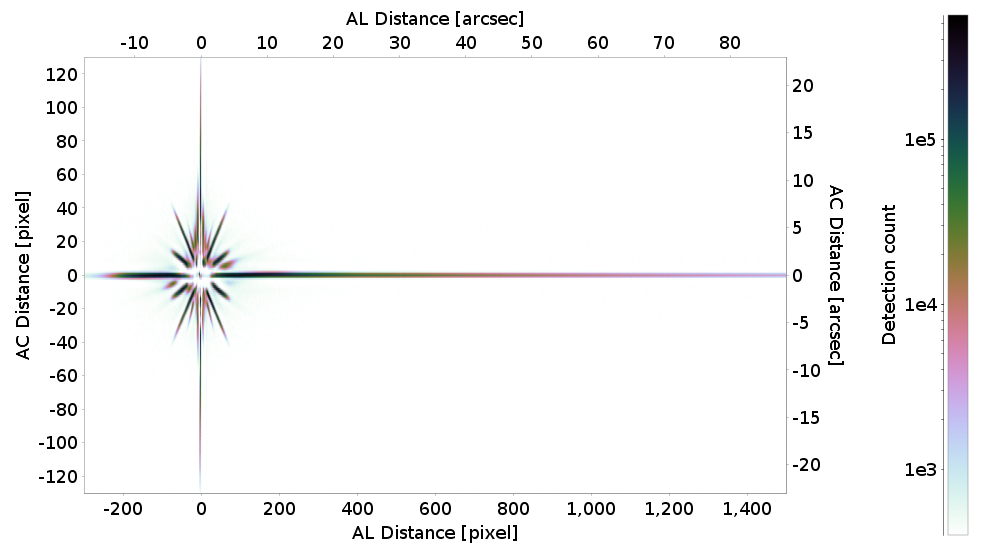}
  \caption{Density map obtained by stacking the onboard detections of about 560 objects in the
  magnitude range  10 to 10.5 during the full \egdr{3} time period. The extent of the diffraction
  spikes, mainly in the AL direction, increases with source brightness. It is larger in the trailing region (to the right)
  as a result of a charge transfer inefficiency in the CCDs.}
\label{fig:dcSm10}
\end{figure}

\begin{figure}[h]
\centering
\includegraphics[width=1.00\columnwidth]{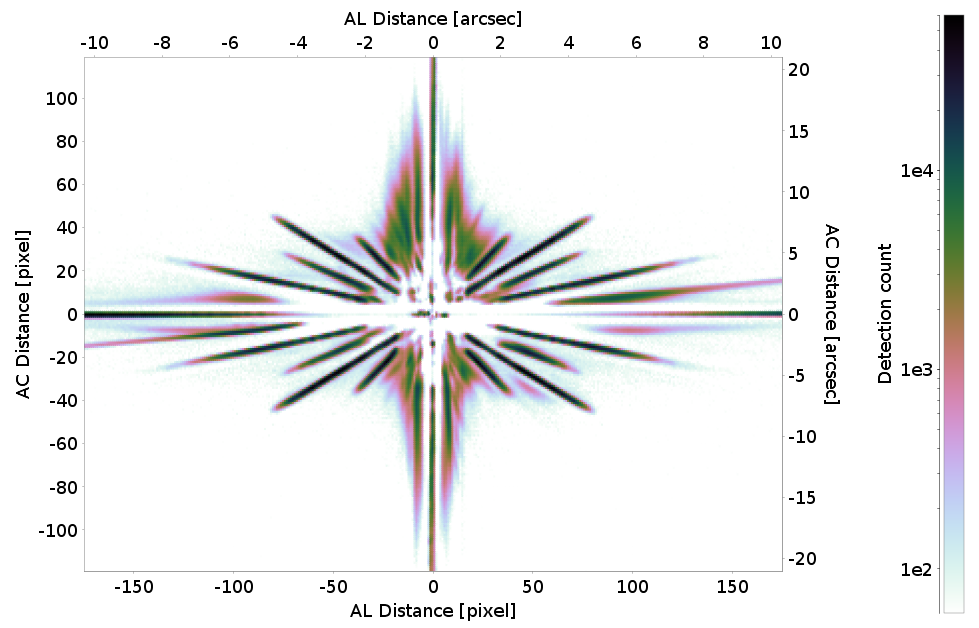}
  \caption{Density map obtained by stacking the onboard detections of about 475 objects
  in the magnitude range  7 to 7.5 during the full \egdr{3} time period. The optical
  diffraction spikes causing the majority of the spurious detections are clearly visible.}
\label{fig:dcSm07}
\end{figure}

Before the XM itself, we therefore classify detections as either genuine or spurious in order to create a list of
false detections. Several categories of spurious detections have been found in the data so far, as
described in \citet{Fabricius2016}, and we consequently need different modules to classify them.
For \egdr{3}, we treat the following cases:

\begin{itemize}
 \item Spurious detections around and along the diffraction spikes of sources brighter than about $G=16$.
 Detections are classified as spurious detections when they fall within a predefined set of regions and
 ranges of magnitude, centred on the bright source. These regions are determined from detection density maps
 such as shown in Figs.~\ref{fig:dcSm10} and \ref{fig:dcSm07}. For very bright objects ($G<6$~mag), we even check
 the detections in the opposite FoV to catch spurious detections coming from unwanted light paths
 in the instrument.
 \item Phantom detections, which are detections considered bright on board ($G<10$~mag), but that lack the expected saturated
 samples in the centre of the SM window. They are caused by confusion
of several bright sources by the onboard detection.
 \item Spurious detections due to cosmic rays, which are rather bright detections in the SM that have % inconsistent flux 
very low signal in the remaining AF windows.
 \item Spurious detections around transits of the major planets. As for the diffraction spikes, we use predefined regions
 and magnitude limits to identify spurious detections.
 \item Detections with odd flux signal profiles in the
 acquired windows, typically from diffraction spikes. This module analyses the window samples and searches for
 meaningful point-like signals within the window core region.
 \item Detections during periods with very noisy attitude solutions and spacecraft attitude issues.
\end{itemize}

The categories for which no countermeasures are in place so far % collect all 
include the spurious detections that occur randomly on the sky and are caused CCD defects and unwanted light paths.
In general, these detections have poor astrometric and photometric parameters and are filtered out in the downstream processes.

\begin{figure}[h]
\centering
\includegraphics[width=1.00\columnwidth]{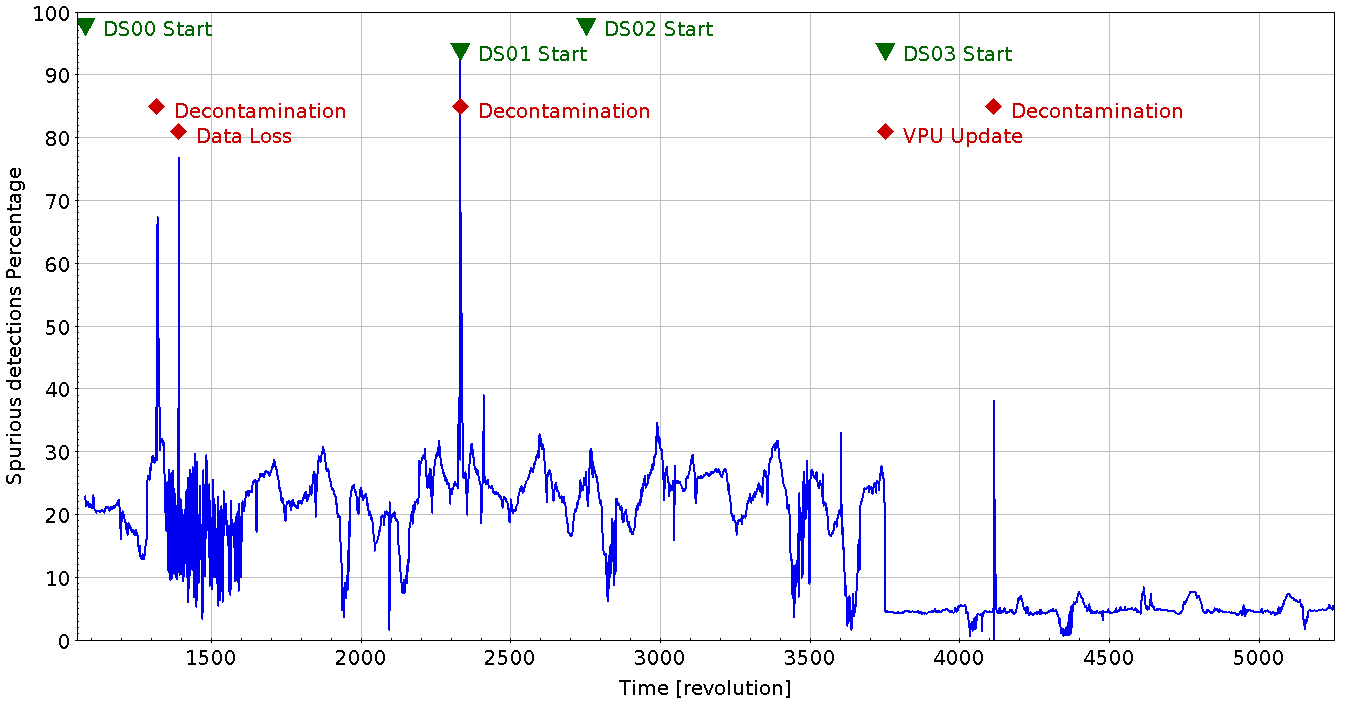}
  \caption{Time evolution of the number of spurious detections in the \egdr{3} data segments. The number of spurious detections
  increases when the spacecraft scans the Galactic plane, as expected, and the major discontinuity at revolution~3750 is related
  to the change in parameters of the onboard detection algorithm (VPU update). 
  The sharp dips are due to interruptions in the detections, and the sharp rises are related to spacecraft events.}
\label{fig:dcEvol}
\end{figure}

For \egdr{3}, the total number of rejected detections is 12\,553~million, corresponding to 16.1\% of all detections.
This is a decrease of 4.6\% in the overall percentage of rejected detections compared to \gdr{2} (from 20.7\% to 16.1\%).
This major percentage update is due to the change of parameters in the onboard detection algorithm that was applied after the closure
of the last data segment entering \gdr{2}. This update was performed to reduce the volume of spurious detections in the
new data segments compared to data segments 0, 1, and 2 (used in \gdr{2}), as shown in Fig.~\ref{fig:dcEvol}.
For data segment 3, the fraction of rejected detections is about 5.1\%.

Moreover, the total number of spurious detections in the data segments 0, 1, and 2 has increased from 10\,737~million
detections for \gdr{2} to 11\,294~million detections for \egdr{3} in the same time. This increase is mainly explained by the new calibration 
used in the module that is in charge of treating the diffraction spikes of bright objects.
On the other hand, the module
analysing the image profiles has also been reviewed in order to avoid rejecting detections of close pairs (separation below 400~mas).
Cycle-02 processing showed indications that this module selected valid detections of close objects when
more than one peak was visible and
the window was not well centred on the brighter peak.
Under these conditions, the module may incorrectly classify some detections that affect the quality of the final astrometric and
photometric solutions. 
We estimate that a non-negligible fraction of these detections remains rejected in \egdr{3,} and furthermore, some
of these close sources may only have a few matched detections.
For this reason, a major update will be carried out in the next processing cycle to include the prompt identification of
these detections and improve their treatment in the XM and subsequent pipelines (see Sect.~\ref{S:conclusions}).

\begin{figure}[h]
\centering
\includegraphics[width=0.99\columnwidth]{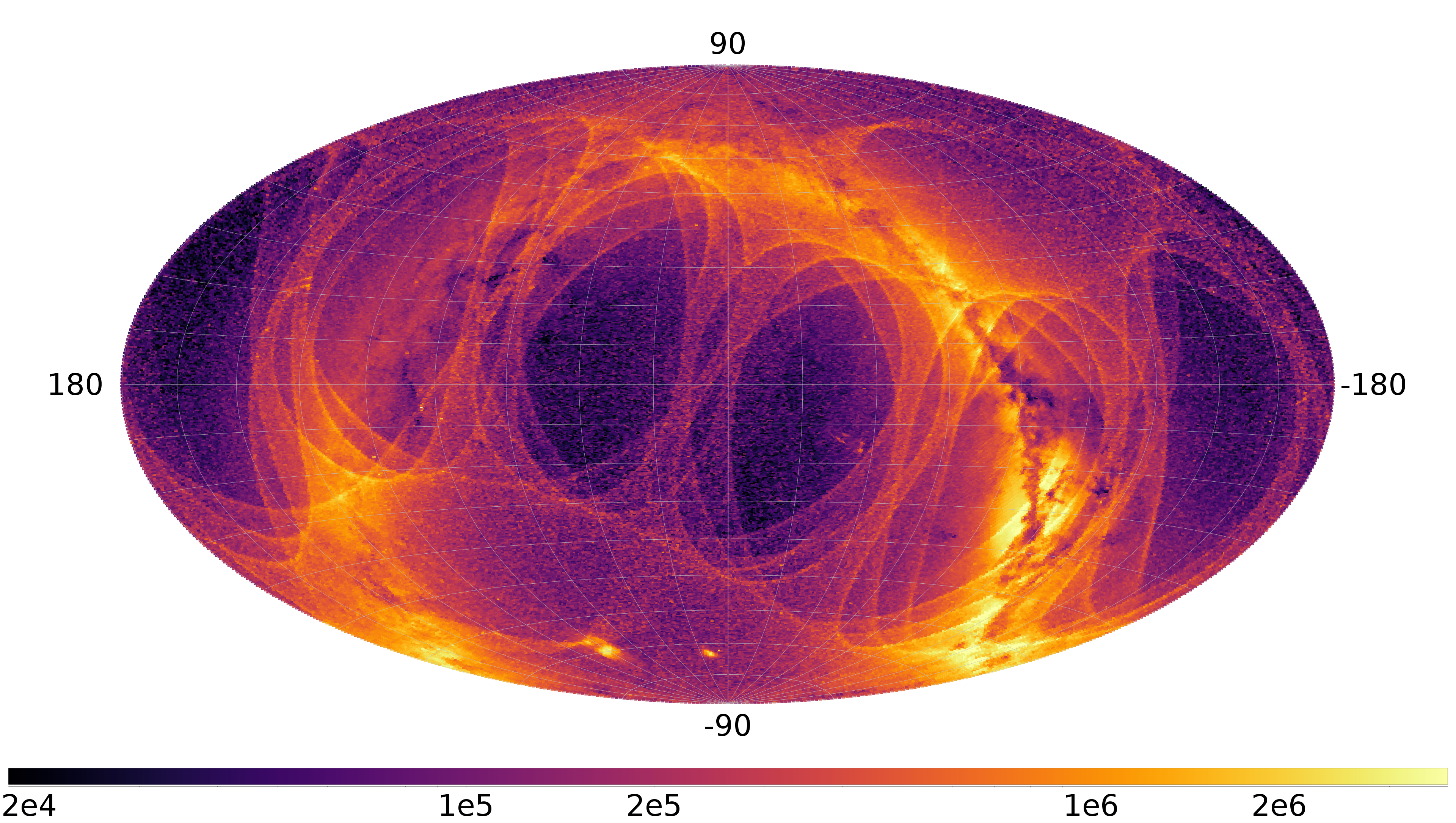}
  \caption{Density sky map (equatorial coordinates) of spurious detections at a pixel resolution of 0.21~deg$^2$.
  The areas with a high spurious detection density form narrow great circles due to the \gaia\ scanning law, where \gaia\ observes more often. The broad circular band is the Galactic plane.}
\label{fig:dcMap}
\end{figure}

\begin{figure}[h]
\centering
\includegraphics[width=0.99\columnwidth]{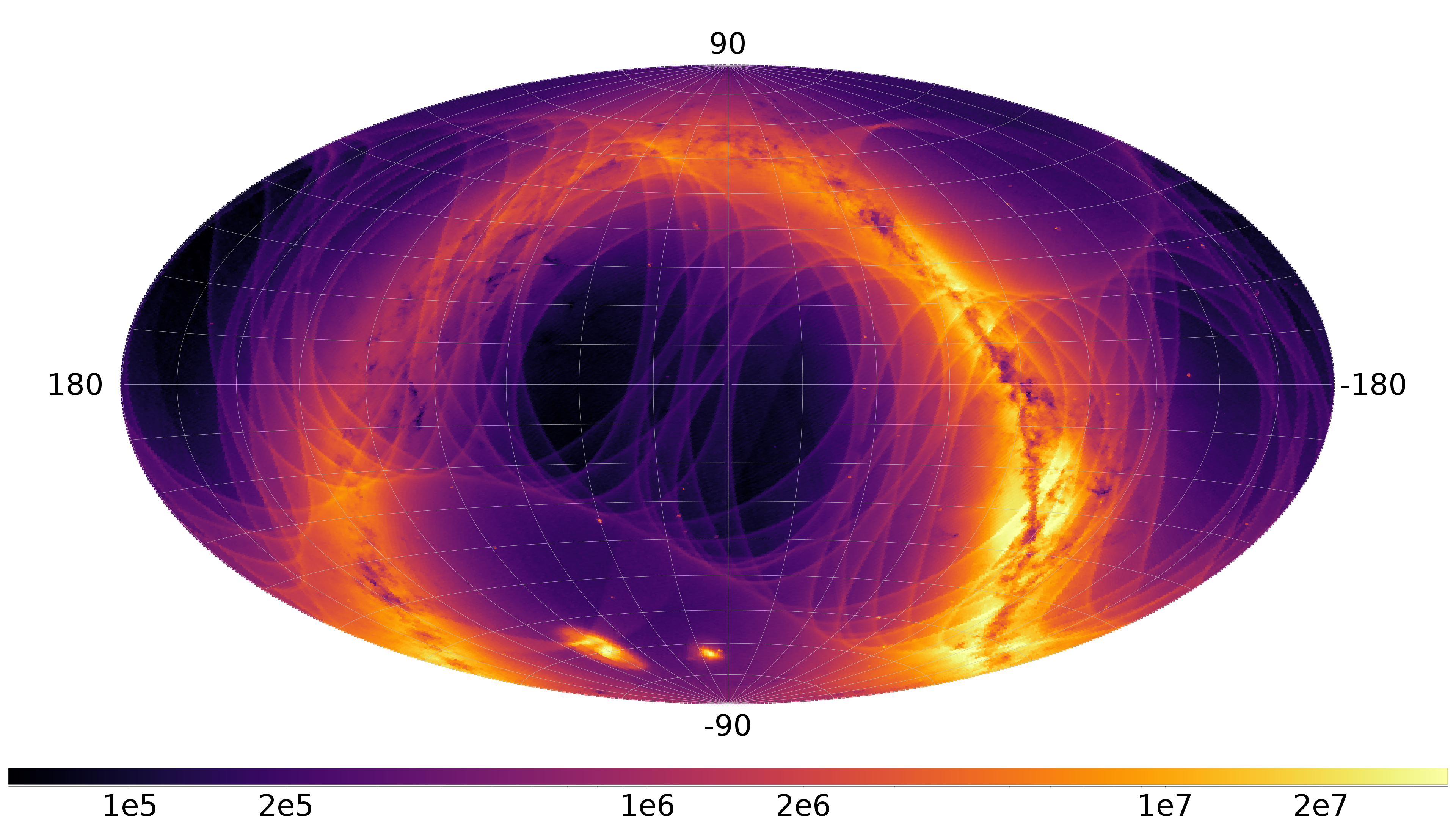}
  \caption{Density map (equatorial coordinates) of detections that are not classified as spurious detections at a pixel resolution of 0.21~deg$^2$.
  The areas with a high detection density are located in dense regions such as the Galactic plane and the Magellanic Clouds. In addition, the sharp circles due to the \gaia\ scanning law are again clearly visible.}
\label{fig:C03McMatched}
\end{figure}

Table~\ref{tb:blByType} specifies the number of rejected detections in each category for \egdr{3}. Most problems come from
the bright sources and the strange image profiles.  
Figure~\ref{fig:dcMap} shows the sky density of spurious detections. It is dominated especially by scanning-law caustics
around the ecliptic plane and in dense areas of the Galactic plane. On the other hand, Fig.~\ref{fig:C03McMatched} shows
the sky density of the detections that enter in the subsequent stages of the XM. In this case, the areas with a
high source density contain more detections, and the great circles due to the \gaia\ scanning law are less prominent.
\begin{table}[h]
\caption{Total number of spurious detections for each category in \egdr{3}.}
\small
\centering
\begin{tabular}{lr}
\hline\hline
\noalign{\smallskip}
Category & \text{Count [million]}\\
\noalign{\smallskip}
\hline
\noalign{\smallskip}
Bright source       & 8159.3 \\
Very bright source  &  158.0 \\
Cosmic and phantom  &    3.7  \\ 
Major planets       &   27.3  \\ 
Odd window profile  & 4066.7 \\
Attitude issues     &  138.8 \\ 
\noalign{\smallskip}
\hline
\noalign{\smallskip}
Total & 12553.8\\
\noalign{\smallskip}
\hline
\end{tabular}
\medskip
\label{tb:blByType}
\end{table}

It may be unavoidable that we classify as spurious some genuine detections that are non-point like sources, such as extended objects. 
For such cases, we accept lists of detections of known extended objects, sources near Solar System Objects, and science alerts provided by
other pipelines, in order to guarantee that these detections 
enter the XM processing unconditionally.

Although the majority of the spurious detections are identified by the above-described criteria, some fraction still remains.
For this reason, an additional module has been included in \egdr{3} to evaluate the quality of the images in the SM and AF
windows in each cluster of detections (see Sect.~\ref{SS:blacklisted}).
This module reviews the classification with a global spatial treatment in order to prevent the
creation of new sources from spurious detections.

Finally, previous to the detection clustering, the surviving detections are analysed by searching for potential double detections.
With \textit{\textup{double detection}} we mean two quasi-simultaneous detections that very likely stem from a spuriously resolved single star. 
These are therefore not spurious detections,
but merely two quasi-simultaneous detections of the same source. If left unmitigated, with both detections entering 
the XM process, a significant quantity of sources brighter than 13 mag would become duplicated in the source catalogue.
Detections from other transits would then be matched to either one or the other source, and we would have an excess of sources.
On the other hand, both detections are necessary for the pipelines dealing with the radial-velocity spectrometer measurements
to be able to reconstruct the full signal that is divided between the two windows in these cases. Therefore these detections are paired before the
clustering, and they are allowed to be matched to the same single source as in previous \gaia~DRs \citep[][Sect.~6.6]{Fabricius2016}. This is in contrast to the default criteria, which force simultaneous
detections to be matched to different sources. In each pair, the fainter detection is flagged as a low-priority detection and is not used  in astrometric or photometric pipelines.

The criterion for identifying these double detections is based on their pixel separation. The threshold for considering them as
double detection depends on magnitude and gradually varies from 13 pixels for $G=5$~mag down to 3 pixels for $G>13$~mag. For \egdr{3}
the threshold for the magnitude range $9<G<13$ has been increased with respect to \gdr{2}. 
This change has implied significant updates on the source list for this magnitude range, reducing the number of duplicated sources,
as described in Sect.~\ref{S:src_assignation}. Moreover, a total of 68 million double detections with low priority was identified in 
this cycle. About 99\% of them are closer than 2 pixels, whereas about 6\% of them are brighter than 13 mag.
About 92\% of these bright detections are closer than 2 pixels.

\section{Determination of isolated groups of detections}\label{S:mcg}

After the classification and cleaning of the input detections, two further steps are
carried out to identify isolated groups of detections and source candidates. Isolated groups are those that can be
processed independently from the other detections and have no source candidates in common between different
groups. We note that a partition of the sky into small isolated groups is essential because of the huge
number of \gaia\ detections; it is not computationally feasible to handle all these data in
a single process.

The first step takes the individual detections seen by \gaia\ in small time intervals from a few minutes to a few hours, according to the detection rate. Then it determines
the observed sky region using the attitude, spacecraft ephemeris, and calibrations and 
retrieves the candidate sources from the working source catalogue, with their positions propagated to
the mean epoch of the input detections.

The source candidates for each individual detection are selected based on their angular distance
to the detection. This distance criterion is the same for all sources and selects all the sources as candidates that are closer to the detection than a given radius.
In future processing cycles, when the working source catalogue is even more precise, we will consider different
criteria for the source candidates, taking advantage of the better precision of sources with
astrometric five-parameter solutions. Moreover, we will consider the better accuracy of the
detection in the AL direction by using an ellipse with the major axis oriented AC.
At this stage, we do not consider the magnitude because we do not wish to discard potential matches to variables sources at this
point in the process.

One of the main objectives of these preparatory steps is that all the detections of a given
source are assembled in the same group even if the source proper motion is not available because of a poor astrometric solution in the working source catalogue.
For this reason, a match candidate radius of 5\arcsec\ was selected that balances the wish to avoid 
creating too large groups on the one hand, and the objective of finding some unknown HPM sources on the other hand.
This value was chosen taking into account that \gaia\ completes a full scan of the sky in six months
and that the fastest moving star in Earth's skies, Barnard's Star, travels 10\farcs3 annually.
We note that if a source is moving faster or if the time gaps between consecutive scans are too long, some of
its observations might be assigned to separated groups of detections. However, if several scans are grouped together, we should be able to determine
a sufficiently good proper motion from the detections in one of the groups, which will enforce the regrouping of all its
detections in the next cycle when the proper motion is available in the catalogue and will be applied when
the candidate sources are applied.

For an impression of the resolution complexity of this first processing step, using a
match candidate radius of 5\arcsec: about 33\% of the detections have only one source candidate in Gaia EDR3, whereas 10\% of them have more than seven source candidates.
The latter is more likely to happen in dense regions, but it may also happen around bright sources where spurious
sources in the working catalogue are common.

\begin{figure}[h]
\centering
\includegraphics[width=0.95\columnwidth]{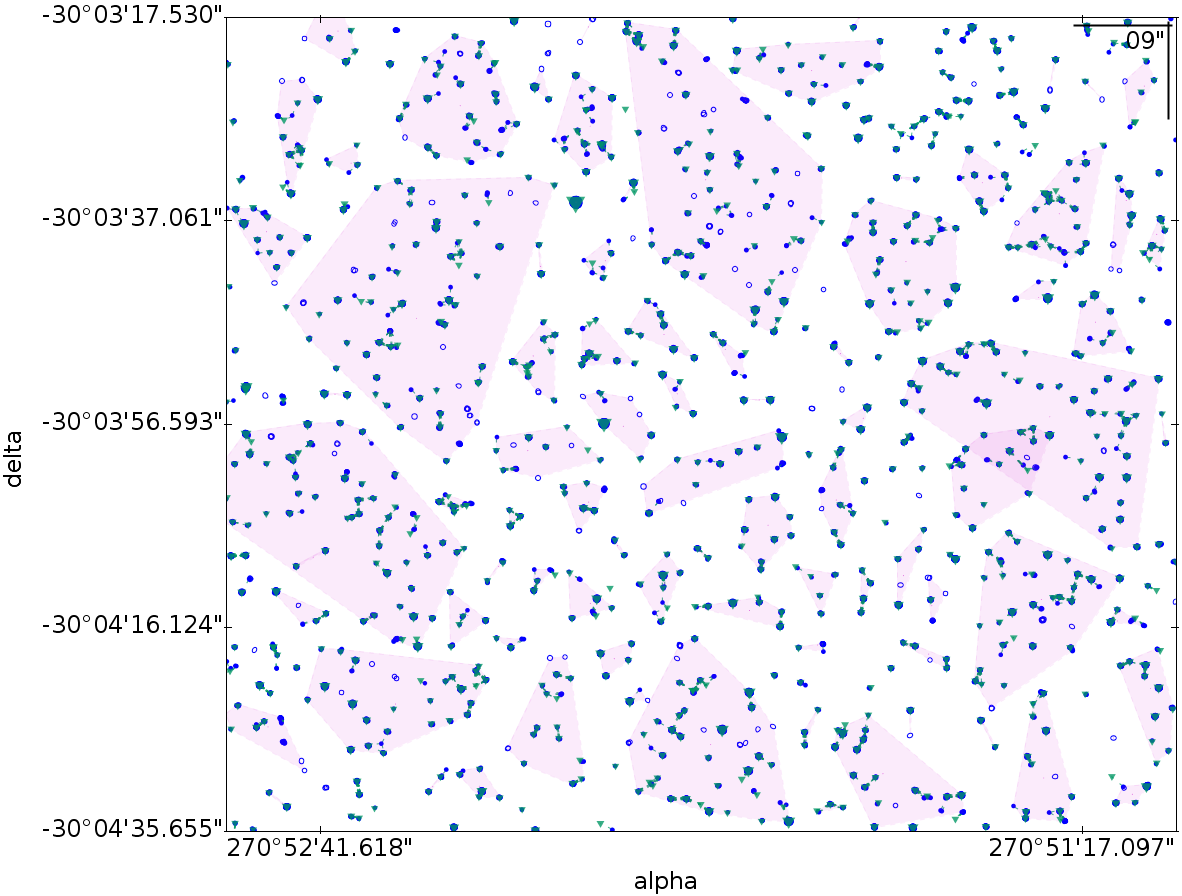}
  \caption{Sample of the isolated groups (pink areas represent their convex hull) of detections (blue dots) and input sources (green triangles) determined
  in a small region of ~2.3 square arcminutes in Baade's Window. If the group is small enough (e.g. a group with only an isolated detection), the pink area is hard to detect in this large area. A total of 10\,400~detections in the \gaia~onboard magnitude range $11.66<G<20.58$ are distributed over 30 scans, and 827 source candidates lie in the \gaia~magnitude range $11.66<G<21.12$. The total of created isolated groups is 145, the largest group has 1554 detections and 126 sources.}
\label{fig:mcgBaadeNgc6522}
\end{figure}

This first processing step produces a set of preliminary assignments of transits to sources that we called
match candidates. In addition, it produces the list of the total number of detections
that are matched to each source.
Each match candidate identifies the potential candidates in the working source catalogue for each detection
and provides the means to assemble the detections in self-contained groups of detections
and source candidates, as shown in Fig.~\ref{fig:mcgBaadeNgc6522}. We note that this is an extreme case with about the highest star density on the whole sky. The majority of isolated groups contain only one input source in most areas of the sky.

These groups are determined in a second processing step that acts as a bridge between the initial time-based
and the final space-based processing \citep{jcPhd}.
This second processing step starts by loading the match candidates for a given sky region.
From the loaded entries, the unique list of matched source candidates is identified and the corresponding global source link counts 
are retrieved. Based on these data, a recursive process is followed to determine the isolated and self-contained groups of 
match candidates and source candidates. In a simplified way, the algorithm proceeds as follows:

\begin{enumerate}[1.]
% Step 1
\item Take the first match candidate and initialise a new group.

% Step 2
\item For each source candidate of the current match candidate,

    \begin{itemize}

    \item check if the number of loaded match candidates linked to the current source candidate matches the global source link count:

        \begin{enumerate}[(a)]

        \item Match candidates are missing, that is, that are not loaded due to the sky region filtering.\\
        The current group is aborted and all collected match candidates are stored for later processing using larger sky regions.\\
        We jump to step 3.

        \item All match candidates have been loaded.\\
        For each match candidate not yet in the group,

            \begin{itemize}

            \item add the match candidate to the group
            \item continue the process recursively from step 2.

            \end{itemize}

        \end{enumerate}

    \item All match candidates of the current source candidate already processed. Check if we have more source candidates:

        \begin{enumerate}[(a)]
     
        \item Take next source candidate and continue from step 2.
        
        \item Close the current group and jump to step 3.
        
        \end{enumerate}

    \end{itemize}

% Step 3
\item Take the next available match candidate (not yet processed), initialise a new group, and repeat the processing from step 2.
\end{enumerate}

As can be deduced from this algorithm logic, the recursive process ends when we have classified all the 
input match candidates into two groups:
\begin{itemize}
\item Closed groups, which are a set of match candidates grouped according to their links to the same group of source candidates. 
This means that all match candidates of all source candidates of a closed group are in that group and are not assigned to other groups.
In this sense, we obtain self-contained and isolated groups of detections and sources.

\item Deferred match candidates, which are a set of match candidates that could not be grouped because further match candidates, linked to the same set of source candidates, are found outside the processed sky region.
\end{itemize}

This process is initially run in parallel jobs following a partitioning of the sky based on the sky density of detections.
In the first run, match candidates are always deferred because the sky distribution of groups of match candidates easily exceed the boundaries of the initial sky regions. This situation is solved by reprocessing all the deferred match candidates with larger sky regions and thus resolving the previous boundary issue.
Sooner or later, a last run consisting of a single job covering the full sky is always needed.

As mentioned above, the match radius we used is large enough to keep detections of high proper
motion stars in the same group of detections. However, this also increments the complexity of the groups,
and occasionally, the groups produced in crowded areas could be too large and complex for the final
clustering resolution.
This complexity naturally increases when more data segments are added. For this reason, a n algorithm to crop the source candidates was implemented for \egdr{3}. This new algorithm reduces the size of the largest groups by discarding
some of the links between match candidates and source candidates, and it recomputes the groups.
Based on a depth-first search technique \citep[e.g.][ ]{Cormen}, the discarded links are
the optimum to reduce the size of the group based on the detection-to-source distances and the total number of
links to the same source candidate. By discarding these links, the original group is split into subgroups
that can be processed later. The links are discarded iteratively until the subgroups contain fewer than a given
number of detections and the angular distance is smaller than a maximum value. These limits are selected
based on the processing performance of the subsequent algorithms, and for \egdr{3,} these limits were fixed to 10000 detections and 1\farcs5.

For \egdr{3}, we created a total of 1198~million isolated groups of detections, and a total of
109~million sources from the working \gaia\ catalogue were discarded because no detections were matched to them.
These sources did not have any detection within the 5\arcsec match radius, probably because they
are spurious sources created in previous cycles from detections that have been rejected in cycle 3.
These results show the importance of the improvements in the detection classifier process to clear the
working catalogue from spurious sources (Sect.~\ref{S:dc}).
About 73\% of the created groups contain only one source candidate, therefore the number of groups with source
ambiguities is small. 
However, 0.3\% of the groups still contain more than 50 source candidates.
We note that the number of groups with one source candidate is much higher than 30\% of individual detections with one source candidate. This is expected because a group with one source candidate only needs the detections of this source candidate, but a group with eight source candidates will need all the detections of these eight source candidates, which could be a large number of detections. Globally, the detections with one source candidate create more groups than the detections with several group candidates.
Ninety percent of the groups with assembled detections contain fewer than 95 detections, which
matches the expectations of the number of scans for isolated single sources.
However, in a few cases, the groups can reach one~million match candidates, for instance for high stellar density regions such as the Galactic centre.

The result of these two preparatory stages is a set of isolated groups of detections, taken from the group of match candidates. This partitioning is
crucial in order to distribute the following processes in a computer cluster and to avoid boundary effects of direct
sky partitioning strategies such as HEALPix. More details of these algorithms are described in \citet{jcPhd}.

\section{Clustering model}\label{S:clustering}

% Reference person for this section: FTC

Cluster analysis aims to divide groups of detections into subsets (called clusters), 
where 
the
detections in each cluster have similar characteristics and are distinct from detections 
within other
clusters.
The clustering model is independent of any existing catalogue (the source candidates we obtained are not used at all here), therefore the input only 
consists of a group of detections. The parameters of the source model are
therefore determined using only the detections in each cluster. 
Following 
this premise, we consider a hierarchical agglomerative algorithm because 
it 
presupposes very little in the way of data characteristics (i.e. in our case, it 
does not require previous knowledge of the number of clusters to be created). 
Moreover, one variant of hierarchical clustering algorithms appears particularly well adapted for high proper 
motions, as suggested by \citet{LL:GAIA-LL-060} in the early stages of the Data Processing and Analysis Consortium (DPAC).

Initially, a cluster is created for each detection. In order to decide which clusters should be merged, a measure of 
dissimilarity between sets of detections is required, which is a positive 
semi-definite symmetric mapping of pairs of clusters onto the real numbers (i.e. $\Delta(C_i,C_j) \geq 0$ and 
$\Delta(C_i.C_j)=\Delta(C_j,C_i)$ for clusters $C_i, \ C_j$). We note that the 
triangular inequality is not necessarily satisfied for our type of problem.

The selected dissimilarity is the Ward method \citep{ward}, which is defined to 
minimise the increase in internal variance. More formally, when $C_i, \ C_j$ are 
two disjoint clusters, the Ward dissimilarity between them is
\begin{equation}\label{eq:ward}
 \Delta(C_i,C_j) = R(C_i \cup C_j)-R(C_i) - R(C_j),
\end{equation}
where $R(C)$ is the sum of squared residuals in cluster $C$.

The sum of squared residuals could in the simplest case be just the angular 
distance between the detections and the cluster centre,
\begin{equation}\label{eq:ssr}
 R(C) = \sum_{O \in C} \sum_{k} w_k(x_k(O)-x_k(C))^2,
\end{equation}
where $x_k(O)$ are the components of the observed data vector $\bm{x}(O)$, 
$x_k(C)$ are the coordinates of the cluster centre vector $\bm{x}(C)$, and 
$w_k$ 
are the weight factors. We would like to point out that the weight factors allow including some coordinates that are not space coordinates, such as the magnitude.

Moreover, the cluster centre, $\bm{x}(C)$, is chosen to minimise the sum of squared residuals. Thus, in the linear case it is 
given by the centre of gravity of the detections in the cluster,
\begin{equation}\label{eq:cluster_center}
 \bm{x}(C) = \frac{1}{n}\sum_{O \in C} \bm{x}(O),
\end{equation}
where $n$ is is the number of detections in cluster $C$.

We write Eq.~(\ref{eq:ssr}) more concisely,
\begin{equation}
 R(C) = \sum_{O \in C} \left\lVert \bm{x}(O)-\bm{x}(C)\right\rVert^2,
\end{equation}
where the weight factors are implied in the norm.

Because our interest is the cluster merging, we write the cluster 
centre of Eq.~(\ref{eq:cluster_center}) in terms of two disjoint clusters 
$C_i$ and $C_j$ such that $C = C_i \cup C_j$,
\begin{equation}\label{eq:cluster_center2}
 \bm{x}(C)=\frac{n_i\bm{x}(C_i)+n_j\bm{x}(C_j)}{n_i+n_j},
\end{equation}
where $n_i$ and $n_j$ are the number of detections in clusters $C_i$ and 
$C_j$, respectively.

In the same manner, using Eq.~(\ref{eq:cluster_center2}), we can rewrite the sum of squared residuals in terms of 
clusters $C_i$ and $C_j$,

\begin{equation}
 R(C) = R(C_i)+R(C_j) +\frac{n_in_j}{n_i+n_j}\left\lVert 
\bm{x}(C_i)-\bm{x}(C_j)\right\rVert^2.
\end{equation}

Therefore the dissimilarity is non-negative definite,
\begin{equation}\label{eq:ward_detailed}
 \Delta(C_i,C_j) = \frac{n_i n_j}{n_i+n_j}\left\lVert \bm{x}(C_i) - \bm{x}(C_j) 
\right\rVert^2.
\end{equation}

This method is selected because the results can be generalised to 
time-dependent models, as discussed below in Sect.~\ref{SS:sourceModel}.

As suggested by \citet{Mullner2011} and \citet{Everitt2009}, an efficient algorithm for hierarchical clustering using the Ward method is the nearest-neighbour chain \citep[e.g.][ ]{murtaghfionn1983,murtaghfionn1987}. This hierarchical algorithm is based on the construction of nearest-neighbour chains and reciprocal nearest neighbours. Specifically, the nearest neighbour of a given cluster $C$ is another cluster, $C'=NN(C)$, such that $\Delta(C,C')$ is the lowest value for every possible cluster $C'$. If $NN(C_i)=C_j$ and $NN(C_j)=C_i$, then $C_i, \ C_j$ are called reciprocal nearest neighbours. Formally, the algorithm builds a chain of nearest neighbours, starting from an arbitrary cluster, until a pair of reciprocal nearest neighbours has been found. These two clusters are then merged, and the chain is modified correspondingly. The process continues iteratively until only a single cluster remains.

The merging of clusters is carried out all the way to the point where 
all detections are in a single cluster, but for the XM process, this makes 
little sense. That is, at the end, all the clusters are merged, but they are merged in a sequence of increasing dissimilarity.
Therefore we consider that the merging only makes sense while the 
dispersion of residuals
within the clusters is below a given limit because it is related to the definition of the Ward method.
This dispersion is measured by the variance $\sigma^2(C) = R(C)/n,$ and the 
limit depends on the error of the input detections (Sect.~\ref{SS:input_obs}) and the source model
error described below in Sect.~\ref{SS:sourceModel}.

Moreover, we add an epoch condition to the algorithm. We consider that two detections are quasi-simultaneous detections when they are detected in the same scan or period of time between two consecutive passes of \gaia~telescopes over a given region of the sky. Because a source can only be observed once per scan and telescope, we include the feature that two clusters cannot be merged when both clusters are considered together in quasi-simultaneous detections. More specifically, according to the separation between the two telescopes and the spin rate of \gaia, the detections in the same cluster have to be separated by at least 106.5~min in time, as explained in \citet[][Sect.~2]{Fabricius2016}.
This condition forces quasi-simultaneous detections into different clusters. It is useful to distinguish detections in crowded areas, as well as close source pairs, where the dissimilarity between detections of different sources may be closer than the dispersion limit. We note that it does not affect the detections that were flagged as double detections in Sect.~\ref{S:dc} because they are paired before this stage.

\subsection{Source model} \label{SS:sourceModel}
 
The Ward method described above assumes that the coordinates 
do not depend on time, which implies that the dissimilarity, Eq.~(\ref{eq:ward_detailed}), depends on the cluster centre vectors and the number 
of detections within the clusters. The cluster centre is the mean of 
the detections, as described in Eq.~(\ref{eq:cluster_center}). However, to include the 
proper motion, we have to consider a linear model for each direction $u$,
\begin{equation}
 u(t)= u_0 + u_1(t - t_0)
 \label{eq:linear}
,\end{equation}
where $u_0$ is the mean position at the mean epoch $t_0$ , and $u_1$ is the proper motion.

The coordinate components are independent of each other in this case, which makes using them in the dissimilarity measure, Eq.~(\ref{eq:ward_detailed}), rather simple and straightforward. We only consider the cases with more than one detection in the cluster, otherwise the time-dependent model is not necessary and we would apply the Ward method described above.

Another source model improvement including the parallax may also be considered, but then the coordinate components are no longer independent. This model is not used in the current processing cycle because it is not regarded as essential, given the precision of the input data and the purpose of the XM (see Sect.~\ref{S:data_used}). However, the source model including parallax may be analysed in future releases when more detections will be added.

The linear system for the determination of $u_0, \ u_1$ in matrix form is 
\begin{equation}
 \bm{b} = \bm{Au} + \bm{e}
 \label{eq:matricial}
,\end{equation}
where $\bm{u} = (u_0, u_1)$,  $\bm{b}$ is an $n-$vector of detections, $\bm{e}$ 
is an $n-$vector of detection errors, and $\bm{A}$ is a $2 \times n-$matrix 
with the time functions.

Thus, the minimum of the sum of squared residuals can be written as
\begin{equation}
 R_u(C)= \bm{b}^T \bm{b} - \bm{\widehat{u}}^T \bm{N} \bm{\widehat{u}},
\end{equation}
where $\bm{N} = \bm{A}^T \bm{A}$ is the normal matrix and $\bm{\widehat{u}} = \bm{N}^{-1} \bm{A}^T \bm{b}$ is the unique solution that minimises the sum of squared residuals. We note that we can guarantee the uniqueness when there is more than one detection in the cluster because then $\bm{A}^T \bm{A}$ is positive-definite and non-singular.

Because our interest is again the cluster merging, we also write $\bm{\widehat{u}}$ in terms of the solutions $\bm{\widehat{u}}_i$ and $\bm{\widehat{u}}_j$ of two disjoint clusters $C_i$ and $C_j$ such that $C = C_i \cup C_j$,
\begin{equation}
 \bm{\widehat{u}} = (\bm{N}_i+\bm{N}_j)^{-1} (\bm{N}_i \bm{\widehat{u}}_i +\bm{N}_j \bm{\widehat{u}}_j),
\end{equation}
where $\bm{N}_i$ and $\bm{N}_j$ are the normal matrices of the clusters $C_i$ and $C_j$ , respectively.
When we apply the initial definition of the Ward method, Eq.~(\ref{eq:ward}),
the dissimilarity in $u$-direction can therefore be expressed as
\begin{equation}
 \Delta_u(C_i, C_j) = (\bm{\widehat{u}}_i-\bm{\widehat{u}}_j)^T 
\bm{N}_i(\bm{N}_i+\bm{N}_j)^{-1}\bm{N}_j(\bm{\widehat{u}}_i-\bm{\widehat{u}}_j).
 \label{eq:diss_pm}
\end{equation}

We would like to point out that this equation reduces to Eq.~(\ref{eq:ward_detailed}) when the normal 
matrices are of dimension $1 \times 1$, that is, without applying the linear model.

Because the coordinates are independent, the minimum of the sum of squared residuals corresponds to the
minimum of the sum of all components. In consequence, the total dissimilarity is the sum of 
the dissimilarities in each coordinate.

However, the proper motion model in the cluster analysis contains a problem in the 
computation of the dissimilarity between two 
clusters of one detection each. In this case, the dissimilarity, Eq.~(\ref{eq:diss_pm}), is always zero, and therefore it would be 
possible to match any two detections 
perfectly. 

The adopted solution is to use the 
zeroth-order source model for the merge 
of clusters with only a few detections, and to use the linear  model, Eq.~(\ref{eq:linear}), when a sufficient number of detections allows a good estimation of the proper motion.

We used the spatial three-dimensional coordinates $(x, y, 
z)$ on the unit sphere. These global coordinates are valid over the whole sphere, and this choice avoids complications that could appear when local or spherical coordinates were used. 

Moreover, we add the mean onboard magnitude difference between clusters as a coordinate. The main reason to consider the mean onboard magnitude difference is that in situations where the ambiguity on the position is high, such as crowded areas, considering the magnitude of the detections significantly improves the clustering performance. 
According to the definition of the Ward dissimilarity, Eq.~(\ref{eq:ward_detailed}), including the magnitude only requires the specification of a scale factor, $w_m$, to make a magnitude error comparable to an error in position. According to the precision of the input data we used (see Sect.~\ref{S:data_used}) and the scientific analysis we performed, the value considered is $w_m=\left(0\farcs5/\text{mag}\right)^2$. That is, the equivalent distance for two detections with one magnitude difference is equal to 0\farcs5. It may affect variable sources of larger amplitudes or sources that have experienced flares. Specific analyses for these cases are therefore discussed below in Sect.~\ref{SS:post-processing}.

The global dissimilarity between two disjoint clusters $C_i$ and $C_j$ accordingly is
\begin{equation}
\Delta(C_i,C_j) = \left\{ \begin{array}{cc}
         \frac{n_i n_j}{n_i+n_j}\left\lVert \bm{x}(C_i) - \bm{x}(C_j) 
\right\rVert^2, & n_i + n_j \leq 3, \\
         \Delta_{\left\{x,y,z\right\} } (C_i,C_j) + w_m \frac{n_i n_j}{n_i+n_j} 
\left(m_i - m_j \right)^2, & n_i + n_j > 3,
\end{array} \right.
\label{eq:global_dissimilarity}
\end{equation}
where $m_i$ and $m_j$ are the mean onboard magnitudes of the detections in clusters $C_i$ and $C_j$, respectively. $\Delta_{\left\{x,y,z\right\}} = \sum_{u \in \{x,y,z \} } \Delta_u$ is considered using the proper motion model from Eq.~(\ref{eq:diss_pm}), and the vector of coordinates includes the magnitude, $\bm{x} = (x,y,z,m)$.

Because the linear model requires more than two detections, we use a threshold of three detections in Eq.~(\ref{eq:global_dissimilarity}). A larger threshold may affect the proper merging of clusters with detections from HPM stars.
The merging is therefore initially carried out in terms of position in order to retrieve a better estimate of 
the proper motion of the source.
 
 As commented above, we considered a stopping rule for the cluster merging according to the source model error and the accuracy of the input detections (Sect.~\ref{SS:input_obs}). It provides a limit for the dispersion of residuals within the cluster, which is measured by the variance $\sigma^2(C) = R(C)/n$. Because this model does not include the parallax, it introduces an error in the model that we have to consider in the dispersion.
 
 We therefore measure the error in the parallax with the cluster amplitude error, $A$, which is calculated as the maximum deviation from the theoretical 
position,
\begin{equation}\label{eq:amplitude}
 A(C)=\max_{O \in C} d( \bm{x}_{\text{pos}}(O) , \bm{x}_{\text{pos}}(C|O)),
\end{equation}
where $\bm{x}_{\text{pos}}(O)$ is the observed coordinates vector, 
$\bm{x}_{\text{pos}}(C|O)$ is the cluster 
coordinate vector at the observation epoch according to the proper 
motion model, and $d$ is the Euclidean 
distance.

The limit for the dispersion of residuals within the cluster is defined as
\begin{equation}
\sigma_{\lim}^2 = \sigma_{\text{pos}}^2 + \sigma_{\text{par}}^2,
\end{equation}
where $\sigma_{\text{pos}} = 0\farcs3$ is the detection error considered according to the precision of the input data used (see Sect.~\ref{S:data_used}),
and $\sigma_{\text{par}}$ is the 
parallax error, which is configured as

\begin{equation}
\sigma_{\text{par}} = \left\{
\begin{array}{cc}
 A(C), & A(C) > \sigma_{\text{t}} \\
 \sigma_{\text{p}_0}, & \text{otherwise,}
\end{array} \right.
\end{equation}
where $\sigma_{\text{p}_0}=0\farcs5$ and 
$\sigma_{\text{t}} = 0\farcs2$. These parameters are configured 
from parallax values of the closest stars. 

Initially, the algorithm uses this high-parallax
value, $\sigma_{\text{p}_0}$, because there is not yet enough information to determine an optimal estimation of the amplitude error. As soon as the amplitude error 
is larger than a threshold, $\sigma_{\text{t}}$, then the maximum cluster error is taken into account instead.

\begin{figure}[h]
\centering
\includegraphics[width=0.49\columnwidth]{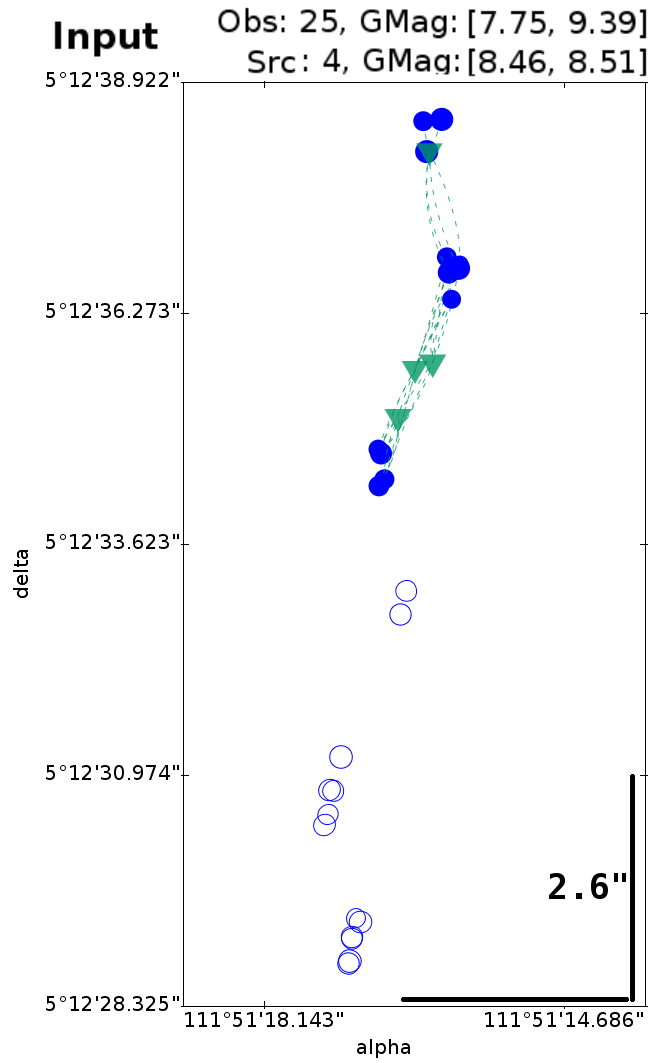}
\includegraphics[width=0.49\columnwidth]{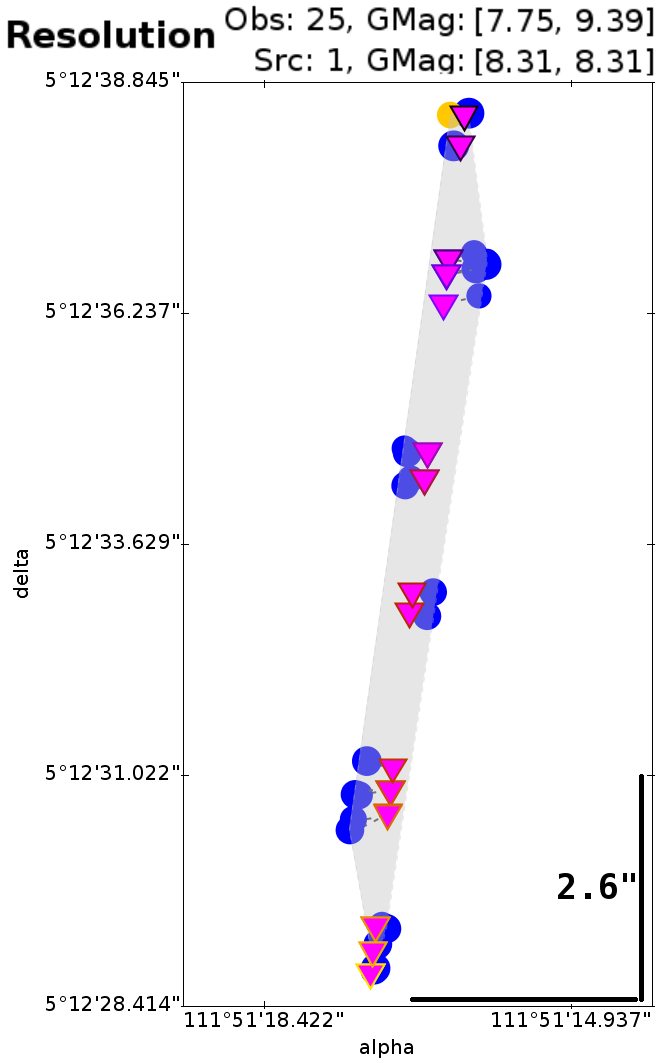}
\caption{XM solution around HIP 36208 (Luyten's star). \textit{Left:} Input candidate sources (green triangles) and isolated group of detections (input of the clustering solution), including those closer than 1\farcs5 to these candidate sources (blue dots) and those without source candidates at a 1\farcs5 distance (blue circles). The dashed green lines are the source candidates links (closer than 1\farcs5). \textit{Right:} XM resolution including the detections (blue dots; the yellow dot denotes double detections with low priority) and the new source propagated to the observation epoch (pink triangles). The grey area indicates the cluster we found. The isolated group of detections in the left panel includes 25 detections in the \gaia~onboard magnitude range $7.75<G<9.39$ and four source candidates in the \gaia~magnitude range $8.46<G<8.51$. The XM solution creates a unique source with a magnitude $G=8.31$ corresponding to the mean of the \gaia~onboard magnitude detection.
\label{fig:hip36208}}
\end{figure}

Figure~\ref{fig:hip36208} shows an example of a HPM source that benefits from the clustering source model. HIP 36208 (Luyten's star) was split into multiple sources with two-parameter astrometric solution in the previous processing cycle. \gdr2 only included the source \gdr2~3139847906304421632 with five matched transits and flagged it as a duplicated source. The XM input for \gaia\ DR3 includes several source candidates that are close to the detections of data segments 0, 1, and 2, whereas the detections of the new data segment 3 do not contain any source candidate closer than 1\farcs5. This radius was used in previous cycles, as explained in \citet{Fabricius2016}, and it is maintained for this and subsequent XM stages in the current cycle to reduce the complexity of the cluster-source assignment (Sect.~\ref{S:src_assignation}). In this case, all the detections are considered in the same isolated group of detections because the match radius of 5\arcsec\ was used when groups in this cycle were determined, as explained in Sect.~\ref{S:mcg}.

The clustering solution for the current cycle assembles all the detections in the same cluster, and then a new source is created with a proper motion that agrees with the expected one. The subsequent pipelines will update the parameters of the new source created using the 25 matched transits. We note that one of these detections is flagged as a double detection with low priority, and it is not used in the clustering model or in the parameters of the created source. This double detection classification (described in Sect.~\ref{S:dc}) reduces the complexity of the clustering solution, and no additional sources are created. The new source created by merging the previous ones, \egdr{3}~3139847906307949696, is released with a proper motion of $\mu_{\alpha^*}=571.23\pm0.04$~mas~yr${^{-1}}$ and $\mu_{\delta}=-3691.49\pm0.04$~mas~yr${^{-1}}$. In addition, the parallax is $\varpi=264.13\pm0.04$~mas. These parameters agree with those published in \citet{Leeuwen2007}, but provide an improvement by a factor of 27 in the proper motion error.

\subsection{Post-processing}\label{SS:post-processing}
In addition to the clustering algorithm, the XM must include a post-analysis of the clustered detections to detect and correct suboptimal cases. This analysis is necessary to validate the clustering solution based on the source model described above. Moreover, this process is needed to modify the suboptimal cases due to spurious detections that we failed to classify correctly, or due to some constraints (and parameters) in the source model that might produce suboptimal solutions in specific cases.
 
The magnitude criterion from Eq.~(\ref{eq:global_dissimilarity}) using the magnitude difference as a coordinate might produce inaccurate solutions. A variable star may create several clusters with different magnitudes at the same position (but from different scans). We therefore have to consider a post-process that detects clusters with very close centres (about
120~mas) and without any common scan (i.e. disjoint in time). After this, they are
merged into a single cluster without any magnitude criterion. Therefore this post-process
prevents the creation of several sources that correspond to a single variable source.

\begin{figure}[h]
\centering
\includegraphics[width=0.85\columnwidth]{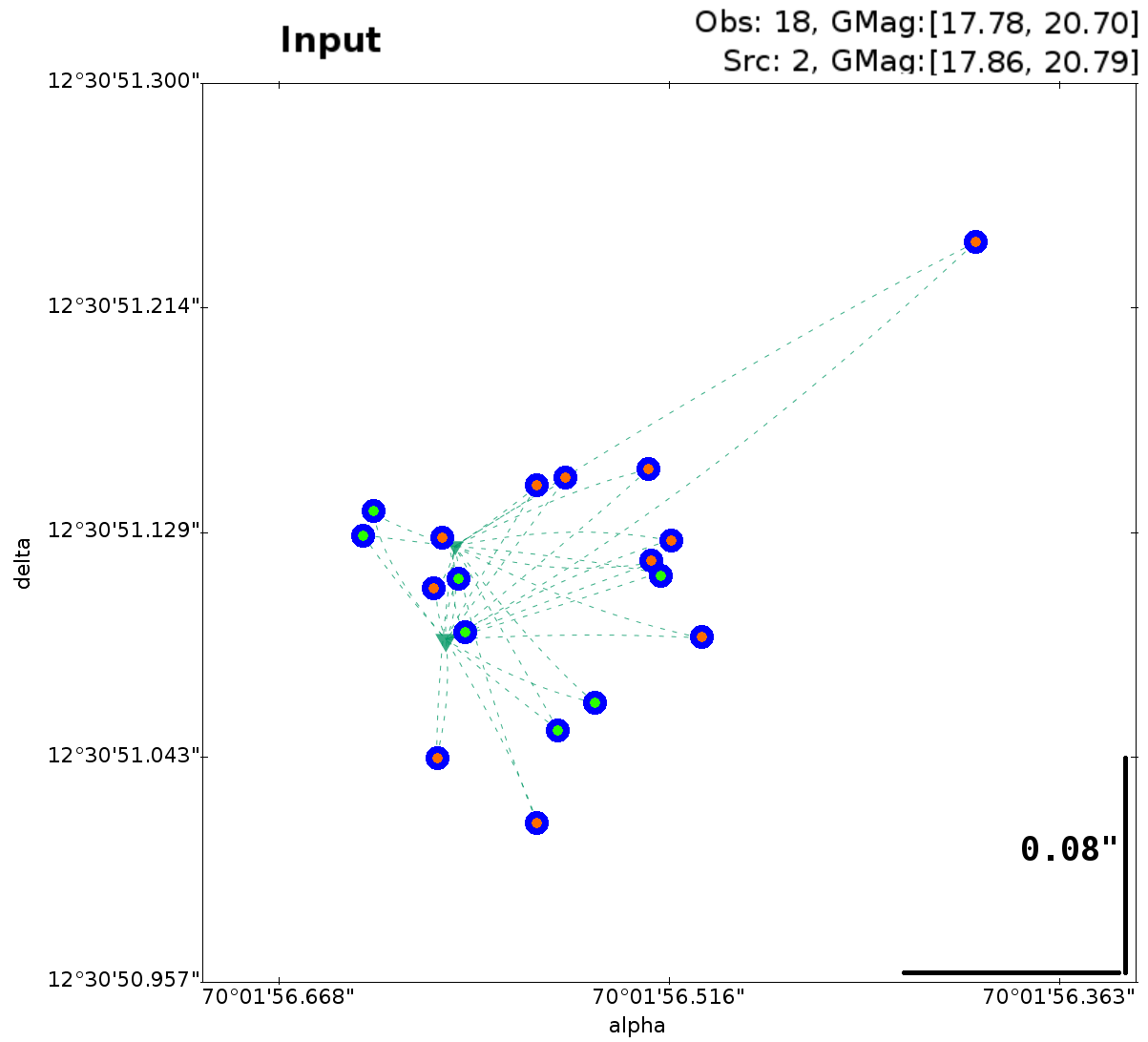} 
\includegraphics[width=0.85\columnwidth]{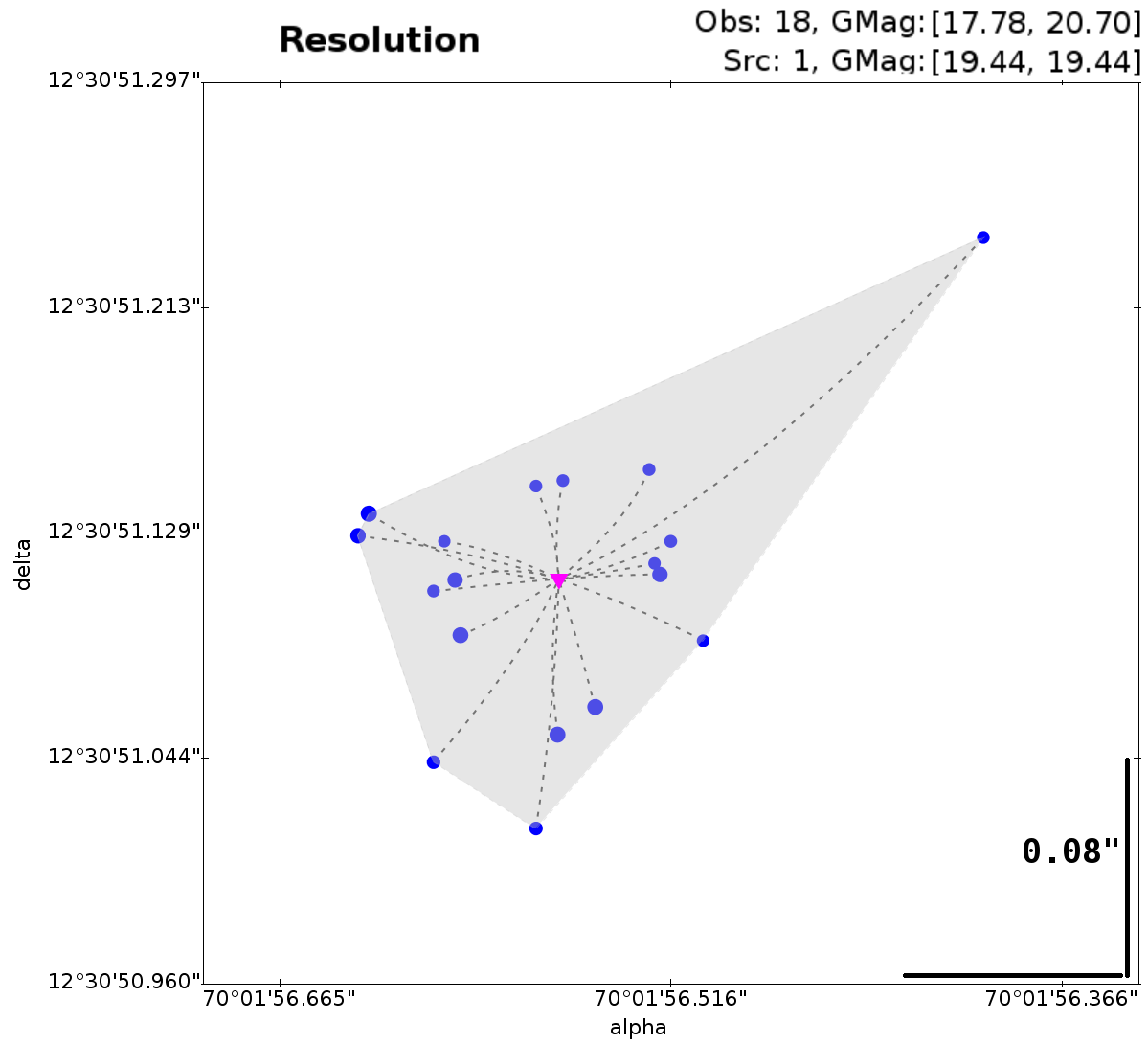}
\caption{XM solution around \gaia~16abo. \textit{Top:} Isolated group of detections (input of the clustering solution). \textit{Bottom:} XM resolution. These plots use the same symbols as in Fig.~\ref{fig:hip36208}. The detections with the green dot are the bright detections matched to one source, whereas the detections with the orange dot are the fainter detections matched to the other source as resolved for \gdr2. The isolated group of detections in the top panel includes 18 detections in the \gaia~onboard magnitude range $17.78<G<20.70$ and two source candidates with a magnitude $G=17.86$ and $G=20.79$. The XM solution creates a unique source with a magnitude $G=19.44$ corresponding to the mean of the \gaia~onboard magnitude detection.
\label{fig:16abo}}
\end{figure}

Figure~\ref{fig:16abo} shows an example of a source that changes its magnitude in an unpredictable way, extracted from \gaia~science alerts \citep{2016arXiv160102827W}.
\gaia~16abo\footnote{\url{http://gsaweb.ast.cam.ac.uk/alerts/alert/Gaia16abo/}} is a spectroscopically confirmed type Ia supernova that shows a declining light curve in agreement with its classification. It was split into two sources in the cyclic processing for \gdr2, the brighter detections being matched with one source, and the fainter detections with the other. \gdr2 contained only the source \gdr2~3307968422513971072 with a two-parameter astrometric solution, based on seven of the brightest detections; it was flagged as a duplicated source. The XM input for the current cycle includes two source candidates from the working catalogue and 18 detections in the same isolated group. The total number of detections in this group has not increased for \egdr3 because the source magnitude extends below the \gaia\ magnitude limit after the DS-02 according to its declining light curve. Because of this post-processing, the XM solution assembles all the detections into one single cluster and creates a unique source \egdr{3}~3307968422514719616 with a magnitude $G=19.44$ corresponding to the mean of the \gaia~onboard magnitude detection. This source is released in \egdr{3} with a low-quality two-parameter astrometric solution based on the few scan directions involved, and a mean magnitude $G=20.44\pm0.06$ determined in the photometric pipeline.

On the other hand, spurious detections may create a HPM cluster, or the detections of a real source may be split into more than one cluster if the dispersion of the detections is larger than the configured parameters described above. In most of these suboptimal cases, the clustering algorithm creates several clusters with a small number of detections each. Thus, all clusters whose number of detections 
are at least half the total number of scans are considered as \textit{\textup{confirmed}} clusters, whereas the other 
clusters are considered as \textit{\textup{tentative}} clusters and have to be analysed and confirmed.

It may be assumed that the sources are observed most of the time when their position on the sky is scanned. Thus, the number of detections in a cluster has to be similar to the number of scans. From this assumption, we derive that the total of detections in the group is similar to the number of clusters times the number of scans in the group. In the same way, we derive that the number of \textit{confirmed} clusters has to be at least the ratio 
of the total number of detections over the number of scans in the group. If this number is not reached, the \textit{\textup{tentative}} cluster with most detections is considered as a \textit{\textup{confirmed}} cluster. This process is iterated until the ratio is reached. 

This assumption ensures that the main part of the solution obtained in the clustering model is not modified in the post-processing. That is, the post-processing analysis does not pretend to create an alternative solution, it is in charge of refining the clustering solution obtained in Sect.~\ref{SS:sourceModel}.

The algorithm therefore analyses whether each detection of a \textit{\textup{tentative}} cluster may be assembled to one of the nearby \textit{\textup{confirmed}}
clusters. The assembly is possible if the following 
conditions hold:
\begin{enumerate}[(i)]
 \item the \textit{\textup{confirmed}} cluster does not contain a quasi-simultaneous detection with regard to any detection in the \textit{\textup{tentative}} cluster,
 \item the variance after the assembly is below the limit described in the source model of Sect.~\ref{SS:sourceModel},
 \item the amplitude error of the assembled cluster is smaller than twice the amplitude error of the \textit{\textup{confirmed}} cluster. More 
formally, when $O_T$ is the detection from the \textit{\textup{tentative}} cluster and $C_{C}$ the \textit{\textup{confirmed}} cluster, then
\begin{equation}
 A(O_T \cup C_C) < 2 \cdot A(C_C),
\end{equation}
where $A(C)$ is the cluster amplitude error defined in Eq.~(\ref{eq:amplitude}).
\end{enumerate}
Because the \textup{\textup{}}confirmed clusters contain a significant number of detections, the amplitude error should not increase substantially when other detections are assembled. Thus, the third condition is crucial in order to prevent this.

Afterwards, some detections from \textit{\textup{tentative}} clusters are not assembled to any \textit{\textup{confirmed}} cluster. These remaining detections are assembled using a clustering process similar to Sect.~\ref{SS:sourceModel}, thus the \textit{\textup{tentative}} clusters are recovered when their detections cannot be assembled to any \textit{\textup{confirmed}} cluster. Then, the created clusters from the remaining detections also become confirmed clusters even when the previous conditions are not fulfilled. We note that the sources at the faint end may contain only a few detections because they are at the detection limit of \gaia. These extreme cases are therefore confirmed and recovered in the last step. Moreover, some of these clusters are created from nearby bright Solar System objects, but a fraction of them is dominated by the remaining spurious detections. A specific spurious-cluster handling is therefore required after this stage, as described in Sect. \ref{SS:blacklisted}.

This post-processing algorithm reduces the constraints of the clustering algorithm and at the same time confirms clusters with only a few detections. Moreover, the algorithm reduces the small number of sources with faulty significant negative and large parallax values listed in \gdr2 even more \citep[see][Appendix C]{Brown2018,Lindegren2018}. 

 \subsection{Cluster classification}\label{SS:blacklisted}
 
The spurious detections that have not been properly filtered out (Sect.~\ref{S:dc}) enter into the clustering stage. In particular, the clusters created mainly from spurious detections might also include some detections that belong to a real source. The post-processing analysis (described above in Sect.~\ref{SS:post-processing}) isolates these spurious detections in small clusters, which in turn may require the creation of new sources.
After the clustering stage, which considers patches of the sky, it is possible to review and improve the classification of the spurious detections with a global spatial treatment. That is, we can classify spurious detections using the cluster information in order to prevent the creation of new sources from spurious detections.

The image parameter determination (IPD) \citep[][Sect.~5]{Fabricius2016} derives the location and flux of the observed
image in the SM CCD and the subsequent nine AF CCDs using the Gaia point and line spread functions \citep{EDR3-DPACP-73}. These parameters together with the XM solution
are then used to refine the astrometric and photometric solution of each individual source in the subsequent pipelines.
However, not all acquired windows can be processed successfully. In certain acquisition circumstances (window truncation, multiple gates, CCD cosmetics, saturation, etc.) the number of samples available in each window is severely limited, causing its automatic rejection during the IPD processing.
We can consider these acquisition circumstances in the XM in order to evaluate the significance of each detection.
In practice, we consider the fraction of usable windows associated with each transit as an indicator of its quality.

Moreover, the estimated error of the sky coordinates of each detection is also considered to classify the detections of each cluster.
Because the accuracy of the attitude is better than 0\farcs1 in general, we use a threshold of 0\farcs5 to identify problematic detections.

For \egdr{3}, we remove a full cluster according to the number of detections. Clusters with a single detection are removed when fewer than five windows can be used, or when the attitude uncertainty is greater than 0\farcs5. Clusters with more than one detection are removed when fewer than three windows can be used, or when the attitude uncertainty is greater than 0\farcs5 in more than half of the detections in the cluster.

Because we consider a similar criterion as in
the IPD, it ensures that the detections that are rejected
in the XM cannot be used in the subsequent pipelines. We therefore achieve first of all a reduction in the number of new sources that are created from scratch and in the number of real sources that are superseded by split because of a spurious nearby cluster.

In the same way as in the detection classification (Sect.~\ref{S:dc}), the cluster classification uses lists of detections provided by other pipelines to guarantee that transits of extended objects, Solar System objects, or science alerts are not removed. A cluster containing such a transit is therefore never rejected.

\begin{figure}[h]
\centering
\includegraphics[width=0.99\columnwidth]{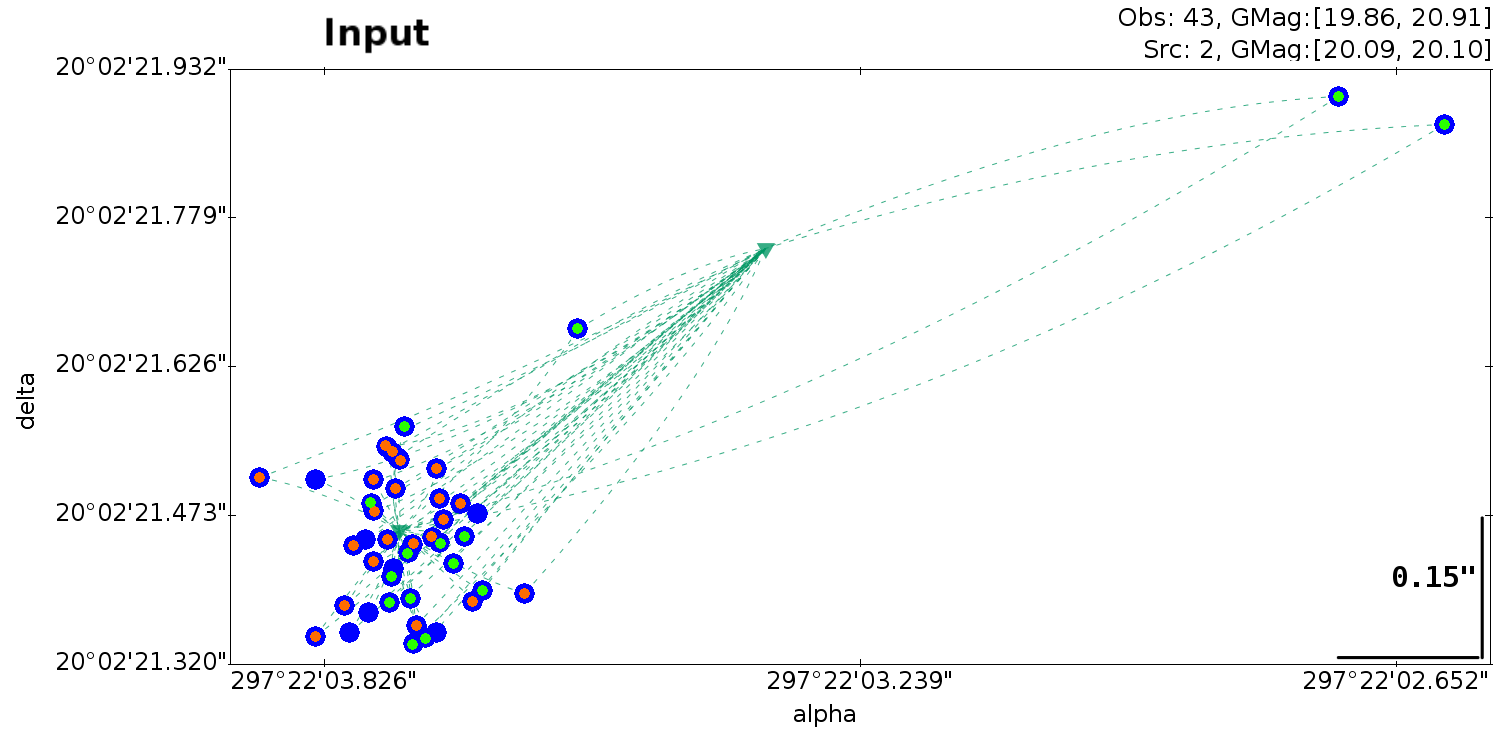} 

\includegraphics[width=0.99\columnwidth]{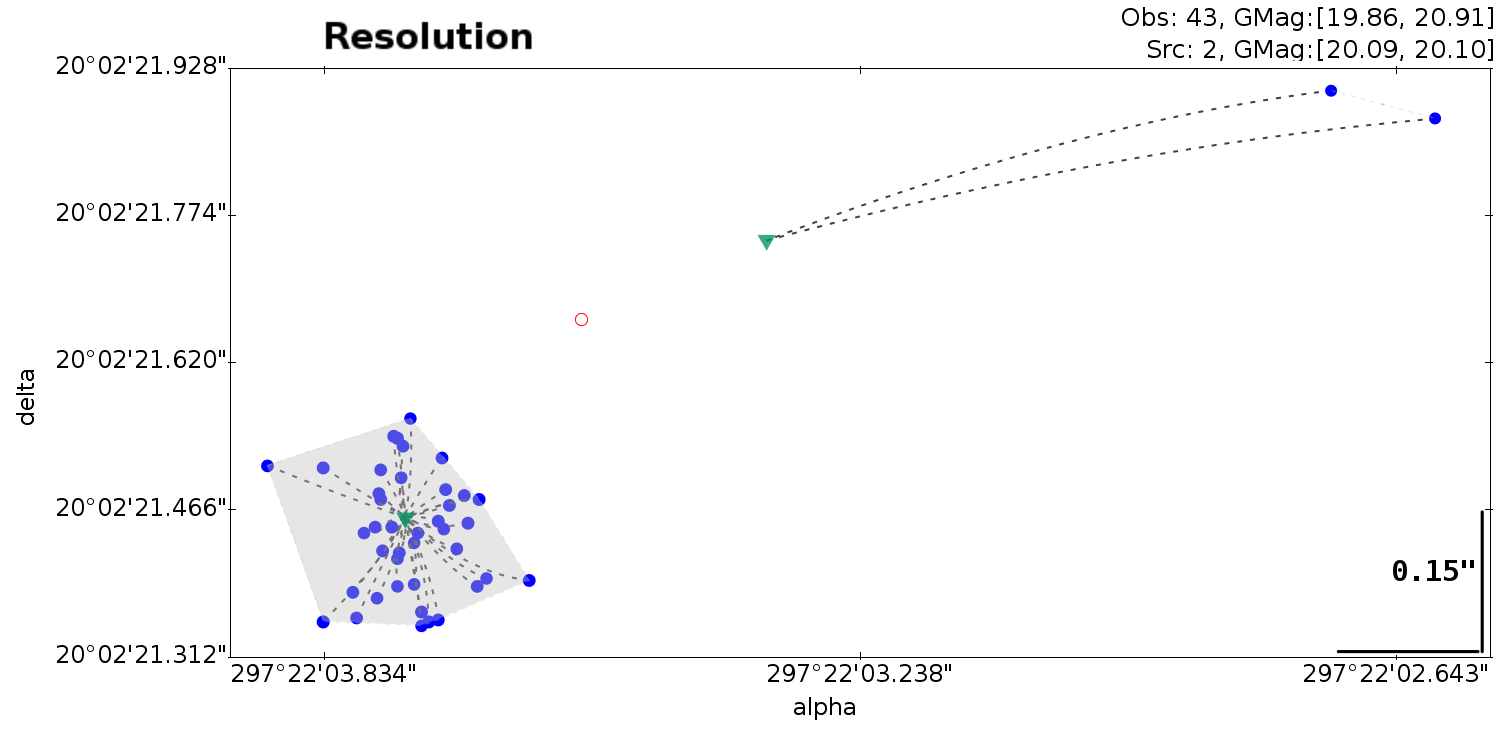}
\caption{XM solution. \textit{Top:} Isolated group of detections (input of the clustering solution). \textit{Bottom:} XM resolution including a rejected detection as an empty red dot. These plots use the same symbols as in Fig.~\ref{fig:hip36208}. The detections with the green dot were matched to one source for \gdr2, the detections with the orange dot were matched to the other source, and the remainder are new detections from DS-03. The isolated group of detections in the top panel includes 43 transits in the \gaia~onboard magnitude range $19.86<G<20.91$ and 2 source candidates with a magnitude $G=20.09$ and $G=20.10$. We note that the positions of already existing input sources are not updated in the XM resolution, they are updated in the subsequent astrometric pipeline run based on the new XM solution.
\label{fig:blcluster}}
\end{figure}

Figure~\ref{fig:blcluster} shows an example of a cluster with an isolated detection that  is rejected at this stage because too few windows can be used. This detection has been isolated in the post-processing analysis (Sect.~\ref{SS:post-processing}) and was rejected afterwards. For \gdr2, these two processes were not available and the clustering algorithm created an inconsistent matching. That is, \gdr2 split the detections of the single real\textit{} source into two different sources: \gdr2~1824801659879180672 with 21 matched transits, and \gdr2~1824801659879180544 with 15 matched transits and a very large parallax ($\varpi=739.3\pm2.6$~mas). However, \egdr{3} includes only one of these sources, \egdr{3}~1824801659879180672, with 40 matched transits, $\varpi = -0.7 \pm 0.5$~mas and $G=20.131\pm0.005$~mag. The other source contains 2 matched transits in the processing for \egdr{3}, therefore  its results are not published.

For \egdr{3}, about 162~million detections have been removed by the classification of about 96~million clusters. About 80\% of the demoted clusters contain a single detection with several discarded windows.

\section{Cluster-source assignment}\label{S:src_assignation}

% Reference person for this section: FTC

As mentioned above, the XM is executed each cycle using the improved source parameters,
spacecraft attitude, instrument calibrations, the updated censoring of spurious detections, and the improved clustering model. Additionally, the clustering algorithm assembles detections without using any information of the input source catalogue. Each new cluster analysis starts from scratch, ignoring any previous match solution, so that an
independent new solution is obtained. The catalogue of the previous release is only used to keep as many source identifiers for the newly created clusters as possible and thus to change the source list as little as possible.

However, the new match solution may update the source identification as well as the astrometric and photometric parameters because new data are added that may change the match in some cases. In addition, some of the
detections will be matched to other sources.

Initially, a list of candidate sources from the catalogue of the previous release is created for each cluster. These are taken from the common subset of the match candidates of all detections of the cluster that was determined when we created the isolated groups (see Sect.~\ref{S:mcg}).

In some cases, the candidate cluster-source link may be unique, but in other cases (especially in crowded areas), multiple candidate cluster-source links may appear which is a contradiction. Therefore, a decision tree algorithm is used to solve all contradictions and to provide the final optimal match solution. For example, two clusters may have the same two source candidates, so that in this case, the decision tree algorithm has to decide which cluster belongs to which source.

In order to reduce the number of candidate cluster-source links, we only consider the source candidates closer than 1\farcs5 at this stage instead of the 5\arcsec considered previously in the creation of the isolated groups of detections (Sect.~\ref{S:mcg}). This radius of 1\farcs5 was configured in the previous \gaia~DRs \citep[see][Sect.~6.5]{Fabricius2016}.

The contradiction of multiple candidate links of each cluster are resolved by breaking the farthest cluster-source candidate link and analysing whether multiple candidate links remain for the current cluster. When no multiple candidate links remain, the next cluster is analysed; otherwise, the second-farthest cluster-source candidate link of the current cluster is removed. This decision is iterated for all clusters until all contradictions are resolved. The possible resolutions of the cluster-source candidate links are listed below.
\begin{itemize}
 \item Assignment of cluster-source: The candidate cluster-source link is isolated, all detections of the cluster are matched to the given source. In this case, the distance cluster-source has to be smaller than 1\arcsec. This value is used to balance the quality of the working catalogue and the wish to reduce assignments to spurious sources. Otherwise, the source is deleted, and a new source is created (see the last two items in this list).
 \item Merge sources: All source candidates of a given cluster are candidates of this cluster alone. In this case, the magnitude difference has to be smaller than 1 mag for all pairs cluster-source. Otherwise, the source is deleted. This condition prevents a \textit{\textup{real}} source from being merged with a potential spurious source.
 \item Split sources: A single-source candidate for several clusters. In this case, each cluster involved has to contain more than about 20\% of the total number of detections linked to the single-source candidate. Otherwise, the source is deleted or assigned to another cluster, if possible. This condition prevents a \textit{\textup{real}} source from being split because of small clusters of spurious detections.
 \item Creation of a new source: The cluster has no source candidates closer than 1\arcsec. This radius is used in order to reduce the assignments to potential spurious sources.
 \item Deletion of a source: The source is not a candidate for any cluster.
\end{itemize}

Each decision affects the options available for further cluster resolution, and this effectively creates a tree 
structure. The order of the analysed clusters determines how the contradictions are solved and therefore the possible final resolution.

In general, we expect the number of multiple candidate cluster-source links to be rather small, therefore it will be possible to explore all permutations. In complex cases, sophisticated approaches have been developed to speed up the search for the optimal resolution and limit unnecessary recursions of
the tree.

To decide which resolution is the the best fit of all the possible resolutions, the following
order of preference was used: 
\begin{enumerate}[(i)]
 \item minimum number of new sources created,
 \item maximum number of sources superseded by merging,
 \item minimum accumulated cluster-to-source distance.
\end{enumerate}
The first condition stabilises the source catalogue and reuses the source identifiers as much as possible. The second condition tries to reduce the number of spurious sources in the catalogue. At the same time, this supplies information of the source evolution for the downstream users (because the superseded sources by merging or split provide the new source identifier to the subsequent pipelines in DPAC, in contrast to deleted sources without matched transits). The third condition usually determines the best-fit resolution based on a pure distance criterion.

The resolution provides the optimal cluster-source links that determine the list of matched transits for each source. The sources are not updated with the cluster parameters determined during the clustering algorithm, although these cluster parameters are used to obtain the best-fit solution. In the case of a new source, there are no previous parameters for the source, thus the source is created using the astrometric cluster parameters and the mean of the onboard magnitude. These parameters are updated in the astrometric and photometric pipelines when possible.

\begin{figure}[h]
\centering
\includegraphics[width=0.94\columnwidth]{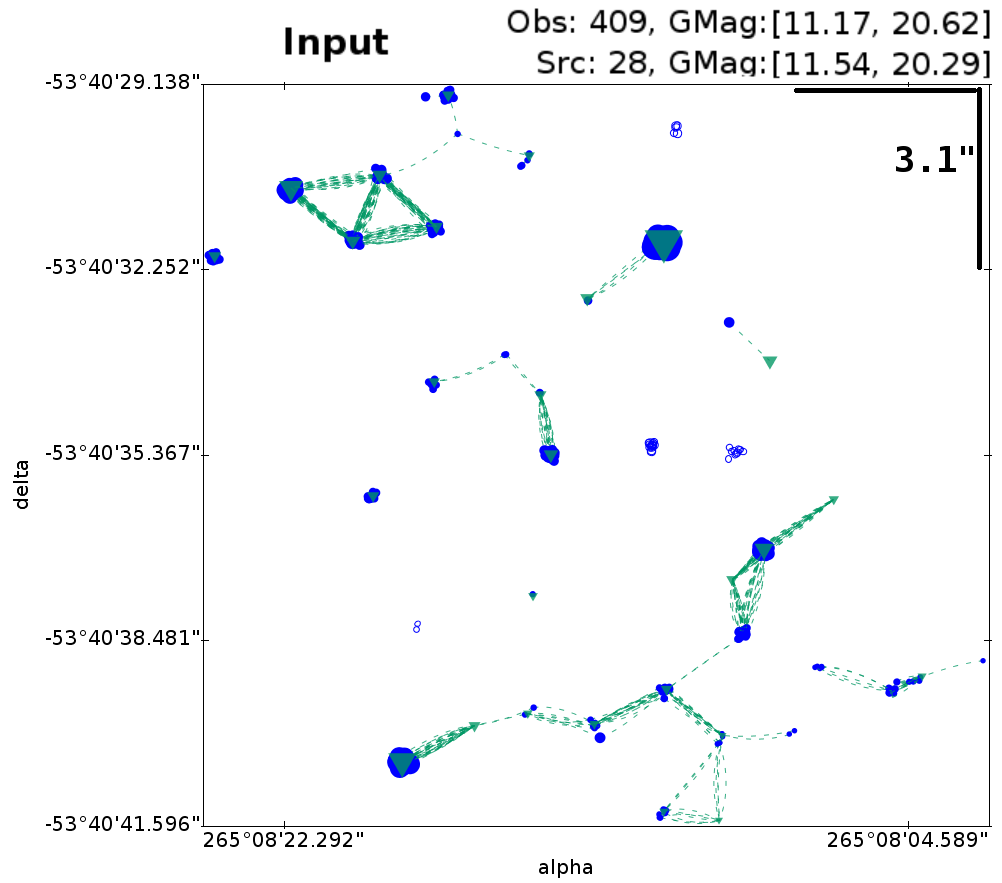} 

\includegraphics[width=0.94\columnwidth]{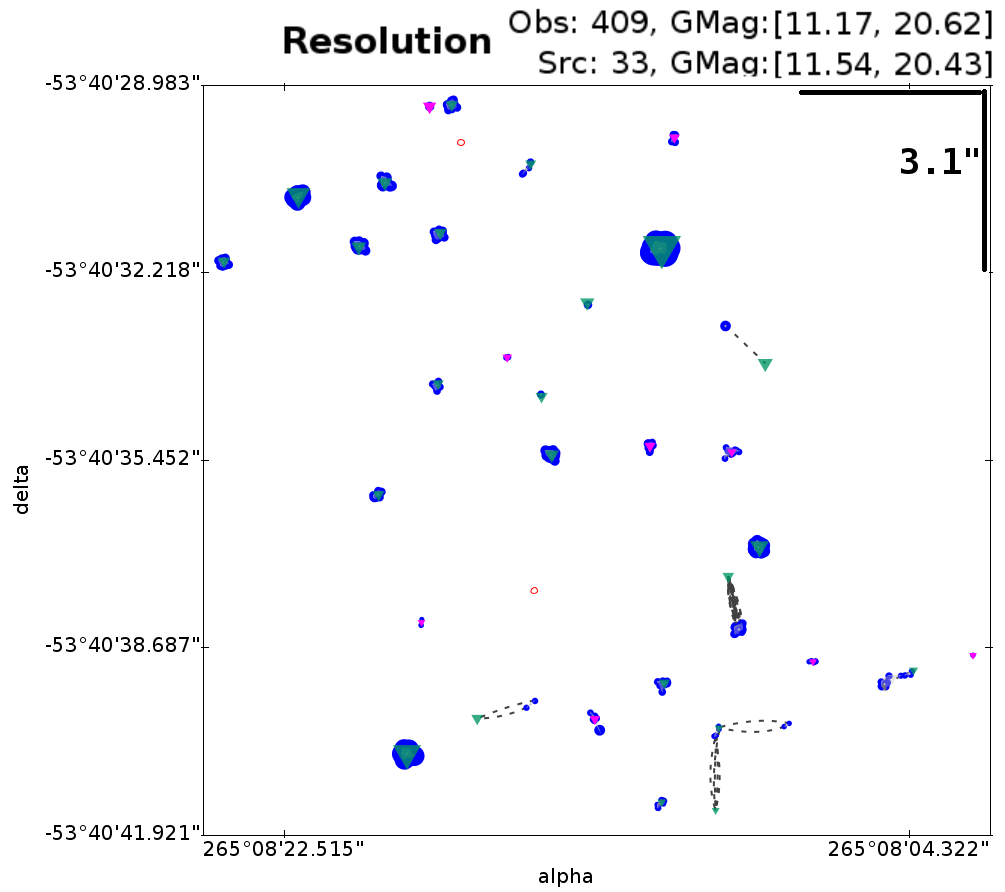}
\caption{XM solution in a dense region. \textit{Top:} Isolated group of detections (input of the clustering solution). \textit{Bottom:} XM resolution including red dots for rejected detections. These plots use the same symbols as in Fig.~\ref{fig:hip36208}. The isolated group of detections includes 409 detections in the \gaia~onboard magnitude range $11.17<G<20.62$ and 28 source candidates in the \gaia~magnitude range $11.54<G<20.29$. The XM solution for the current cycle contains 33 sources in the \gaia~magnitude range $11.54<G<20.43$.
\label{fig:dense}}
\end{figure}

Dense regions may include groups of hundreds of thousands of detections, which may cause several multiple cluster-source links. 
Figure~\ref{fig:dense} shows a small region with several multiple candidate cluster-source links that can be analysed visually. In this case, the XM solution creates 8 new sources from scratch, and only one created by merging. On the other hand, 24 of the 28 input sources persist, 2 are superseded due to a merger, and 2 are deleted because their detections of the previous cycle having been classified as spurious detections.

\section{Results and validation}\label{S:validation}

% Reference person for this section: FTC

This section summarises the scientific results of specific analyses in order to validate the XM solution. The aim is to characterise the source and transit evolution in the XM solution and the improvements for the HPM sources.

The analyses of this section report the behaviour of the source list after the XM solution, which is the input source list for the remaining pipelines in DPAC for \egdr3. The statistics for the published \egdr3 source list are described in \citet{EDR3-DPACP-126}.

\subsection{Source evolution}\label{SS:src_evolution}

For \egdr{3}, a total of 65\,255~million detections are matched to 2552~million sources, which is a reduction of about 30~million sources compared to the input catalogue (Sect.~\ref{SS:src_catalogue}). A total of 2286~million sources (89\%) persist from the input catalogue in the output of the XM, whereas 25~million sources were created by split, 45~million sources were created by merging, and 194~million sources were created from scratch. 

The source evolution in terms of merged and split sources relative to the input source catalogue (Sect.~\ref{SS:src_catalogue}) is shown in Table 2 for different magnitude ranges. Most of the sources still exist after the XM assignment, although about 7.3\% of the sources are deleted. Of these 7.3\% of sources without matched transits, about 57\% have been deleted because they did correspond to a spurious detection that was filtered out in this cycle. The remaining 43\% of the sources without matches are deleted during the final cluster-source assignment (Sect.~\ref{S:src_assignation}) because some sources cannot be merged or split due to the criteria explained above. We note that the percentage of deleted sources is higher at the faint end because spurious detections are more frequent in this magnitude regime. On the other hand, the percentage of sources that are superseded by merging increases for the brighter sources because the double-detection threshold is recalibrated, as described in Sect.~\ref{S:dc}.

\begin{table*}[h]
\caption{Source number evolution of the input source catalogue (Sect.~\ref{SS:src_catalogue}) for different magnitude ranges. We note that these are accumulated values.}
\small
\centering
  \begin{tabular}{cS[table-format=5.2]rrrr}
  \hline\hline
  \noalign{\smallskip}
  Magnitude & \text{Input sources [million]} & Merged & Split & w/o matches & Persisting \\ 
  \noalign{\smallskip}
\hline
\noalign{\smallskip}
  $G < 8$ & 0.086 & 6.3\% & \textless0.1\% & 15.7\% & 77.9\% \\ 
$G < 11$ & ~1.58 & 15.7\% & \textless0.1\% & 3.3\% & 80.9\% \\ 
$G < 14$ & ~19.21 & 4.7\% & 1.7\% & 1.5\% & 92.1\% \\ 
$G < 17$ & ~181.75 & 3.3\% & 3.6\% & 2.3\% & 90.8\% \\ 
\noalign{\smallskip}
Full & ~2582.61 & 3.6\% & 0.4\% & 7.3\% & 88.7\% \\ 
\noalign{\smallskip}
\hline
  \end{tabular}
  \medskip
  \label{tb:MDB02magnitudeRange}
\end{table*}

\begin{figure}[h]
  \begin{center}
  \includegraphics[width=0.95\columnwidth]{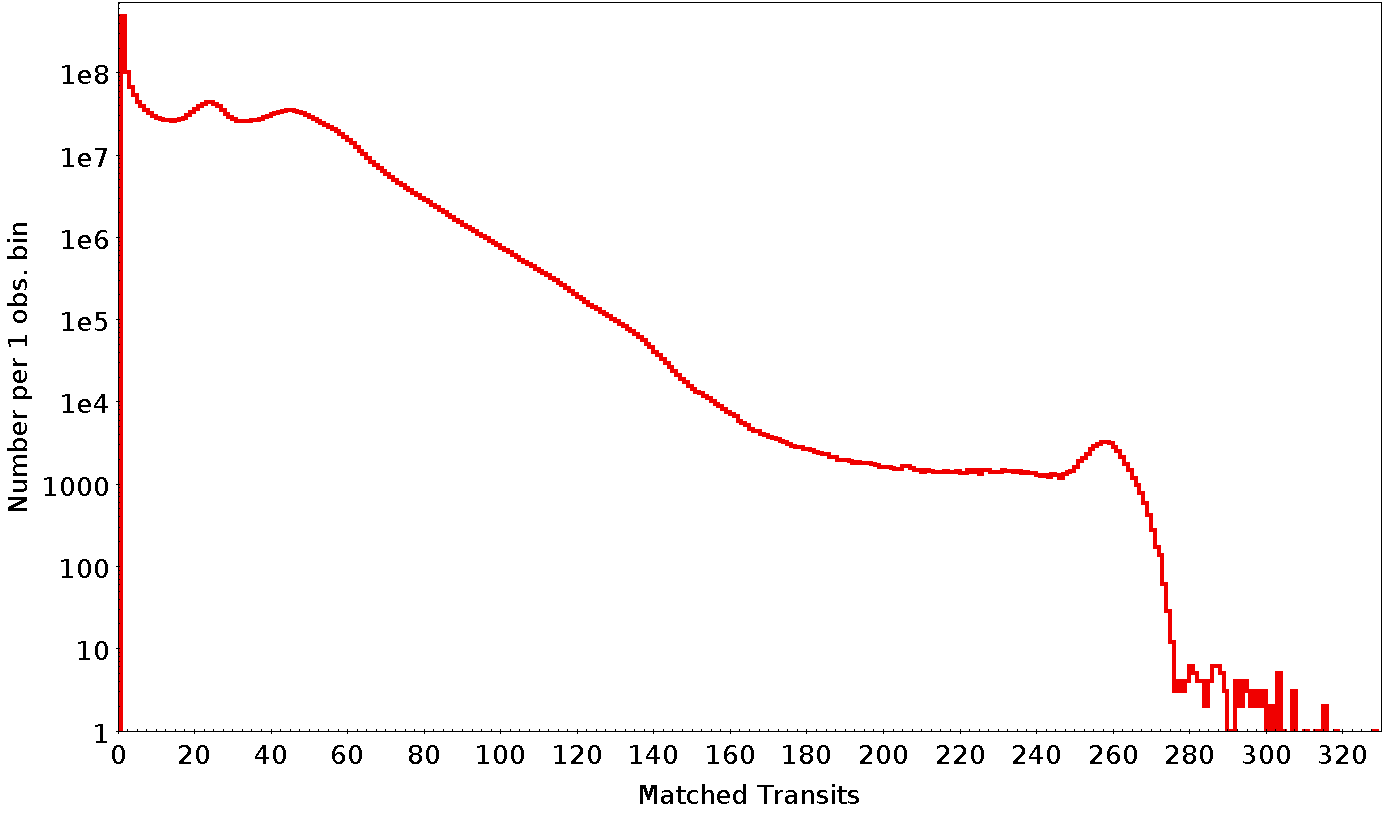}
  \caption{Distribution of the number of matched transits per source after the XM solution for \egdr3.}
  \label{fig:C03IntegratedSourceMatches}
  \end{center}
\end{figure}

Figure~\ref{fig:C03IntegratedSourceMatches} shows the distribution of the sources after the XM solution as a function of the number of transits matched to them. About 60\% of the sources have between 10 and 67 matched transits. One percent of the sources have at least 87 matched transits. The peak of sources with only one matched transit suggests that some spurious detections remain, as well as some sources at the faint end that cannot be detected regularly. This holds, for example, for about 91\% of the new sources created from scratch. Some of these sources with just one match are Solar System objects, but they are mostly spurious and no parameters are published for them. Most of them are new sources created from scratch, but some spurious detections may be assigned to an existing source as well. We note that not all of these sources make it to the release because this depends on the astrometric and photometric filters \citep{EDR3-DPACP-128,EDR3-DPACP-117}. Nevertheless, about 1800~million sources have at least 5 matched transits.

Because spurious detections are treated differently in processing cycle 3 compared to \gdr2, the evolution of the published \gdr2 source list may also be interesting. Of the total of 1693~million published sources in \gdr2, 1651~million sources remain after the XM solution for \egdr{3}. Table~\ref{tb:gdr2_persisting} shows the distribution of surviving \gdr2 source identifiers for different magnitude ranges.

\begin{table}[h]
  \caption{Source evolution of the \gdr2 sources for different magnitude ranges after the XM solution.}
\small
  \centering
  \begin{tabular}{cc}
  \hline\hline
  \noalign{\smallskip}
  Magnitude     & \gdr2 persisting \\ 
  \noalign{\smallskip}
 \hline
   \noalign{\smallskip}
    $G<5$       & 59.0\% \\
    $5<G<8$     & 94.8\% \\ 
    $8<G<11$    & 90.2\% \\ 
    $11<G<14$   & 97.9\% \\ 
    $14<G<17$   & 98.3\% \\ 
        $17<G$  & 97.4\% \\ 
        \noalign{\smallskip}
    Full        & 97.5\% \\ 
    \noalign{\smallskip}
\hline
  \end{tabular}
  \medskip
  \label{tb:gdr2_persisting}
\end{table}

\begin{figure}[h]
  \begin{center}
  \includegraphics[width=0.95\columnwidth]{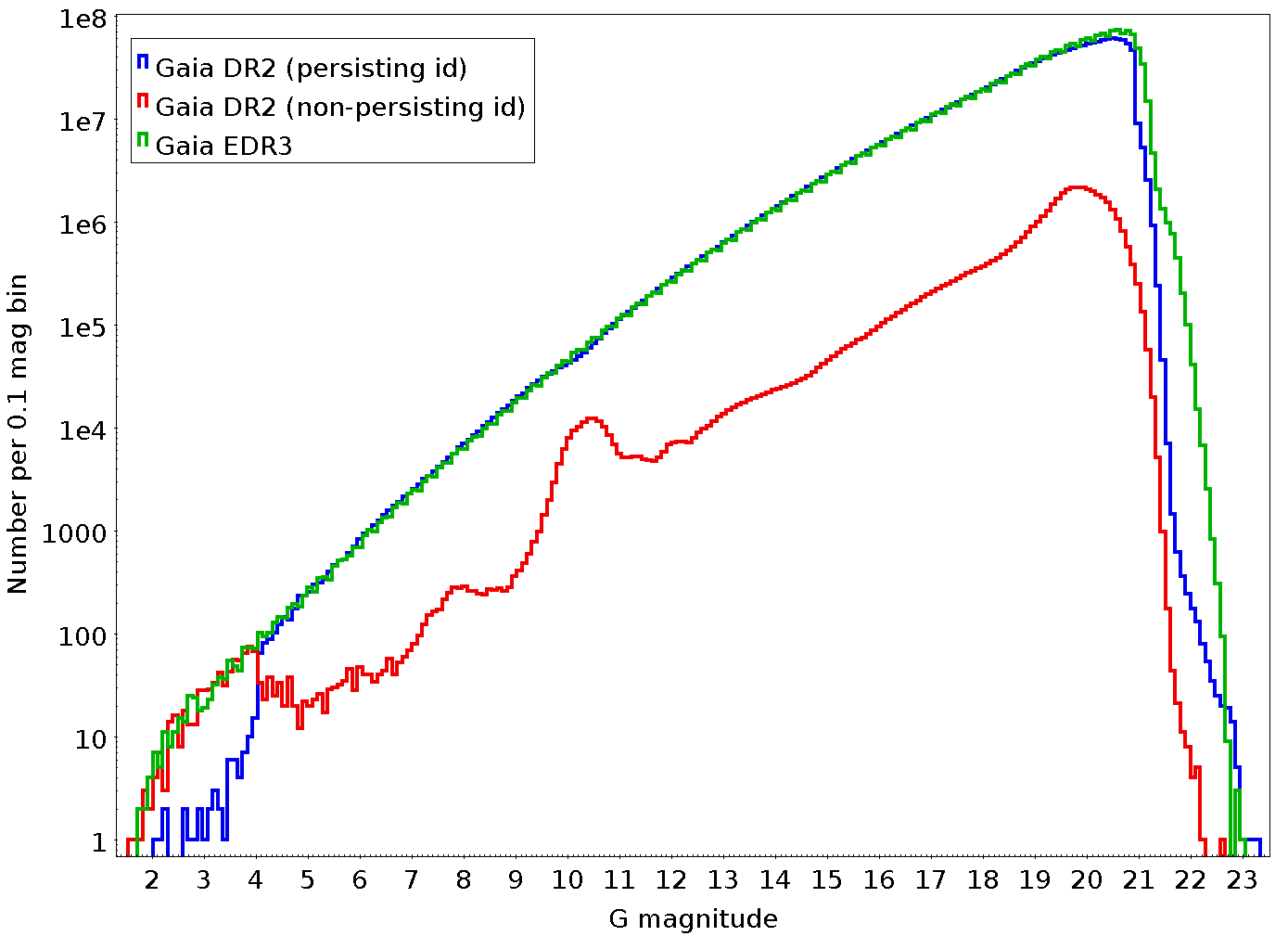}
  \caption{Distribution of the published \egdr{3} sources (in green) as a function of magnitude compared
to the distribution of the published \gdr2 sources that retain their source identification after the XM solution for \egdr{3} (in blue) or change it (in red).}
  \label{fig:C03Gdr02SupersededMagnitude}
  \end{center}
\end{figure}

The bump between the magnitude range $9.5 < G < 11$ shown in Fig.~\ref{fig:C03Gdr02SupersededMagnitude} contains about 0.12~million sources, 13\% of the total number of \gdr2 sources in this range. It is related to a higher occurrence of double detections for the saturated sky-mapper images in this magnitude range. In this cycle, most of these sources are superseded by merging because the double-detection threshold is recalibrated, as described in Sect.~\ref{S:dc}, which also implies a reduction of the duplicated sources in this range.

Several very bright sources ($G<5$) have a close-by source at a distance shorter than 1\farcs5 in the input source catalogue with a similar magnitude, so that most of them take part in a cluster-source resolution stage. Specifically, about 45\% of the very bright sources in \gdr2 were duplicated. Because of that, the fraction of sources that are superseded by merging or that are deleted (therefore no longer persist) in this magnitude range is large, as shown in Fig.~\ref{fig:C03Gdr02SupersededMagnitude}. Because most of their detections are not filtered out, \egdr{3} includes these very bright sources as new sources.

This comparison with respect to the \gdr2 catalogue includes all the sources that are matched in the XM solution. Although the source evolution is mainly driven by the XM process, the total number of \gdr2 sources that are released as \egdr{3} sources also depends on the parameters that can be derived in the astrometric and photometric pipelines \citep{EDR3-DPACP-128,EDR3-DPACP-117}, and also on the internal validation \citep{EDR3-DPACP-126}.

\subsection{Transit evolution}

There is a total number of 65\,255~million matched transits. Most of the detections have been matched to previously existing sources in the input catalogue (96.7\%), and the match distance for about 90\% of the transits is smaller than 200~mas. Larger distances may indicate that the source is reused and that its astrometric parameters have been updated for \egdr{3}.
Although about 96.7\% of the detections are matched to an existing source, about 11\% of the sources are new. The reason for this is that a large number of new sources contain only one matched transit, as explained above in Sect.~\ref{SS:src_evolution}. Moreover, these isolated detections are observed during the new data segment for about 72.5\% of the cases. 

About 86.3\% of the matches are unambiguous (i.e. any other source closer than 1\farcs5), 11.9\% have one ambiguous source, and the remaining 1.8\% have more than one ambiguous source. This indicates that most of the cluster-source links are unique, so that this ensures the stabilisation of the main part of the source catalogue.

The cluster-source assignment discussed in Sect.~\ref{S:src_assignation} takes into account the astrometric and photometric parameters of the input sources and the clusters obtained previously in the clustering algorithm (Sect.~\ref{S:clustering}). Because it does not consider the XM solution of previous releases, the transits matched to a given source may evolve between \gdr2 and \egdr{3} for objects that have retained their source identification. 

In the source list after the XM solution in \egdr3, about 84\% of the input sources retain all their matched transits from the previous XM solution. About 62\% of them accumulate additional matched transits from data segment 3 alone, but about 19\% of them obtain additional matched transits from the period of time used in the previous release. The remaining 19\% do not accumulate any additional matched transit mainly because they only contain one matched transit (for about 78\% of these remaining sources).

On the other hand, the percentage of sources after the XM solution that retain less than 50\% of their matched transits in the input catalogue is 1.55\%, whereas only 0.95\% of the sources lose all their matched transits from the \gdr2 XM solution. Typically, two close sources may swap their linked transits fully or partly. This may also occur around spurious sources and in dense regions. We note that in these cases the source identifiers switch from one physical source to another, thus their astrometric and photometric parameters are updated for \egdr3 as well as the rest of their source parameters for \gdr{3}. 

\begin{figure}[h]
  \begin{center}
  \includegraphics[width=0.95\columnwidth]{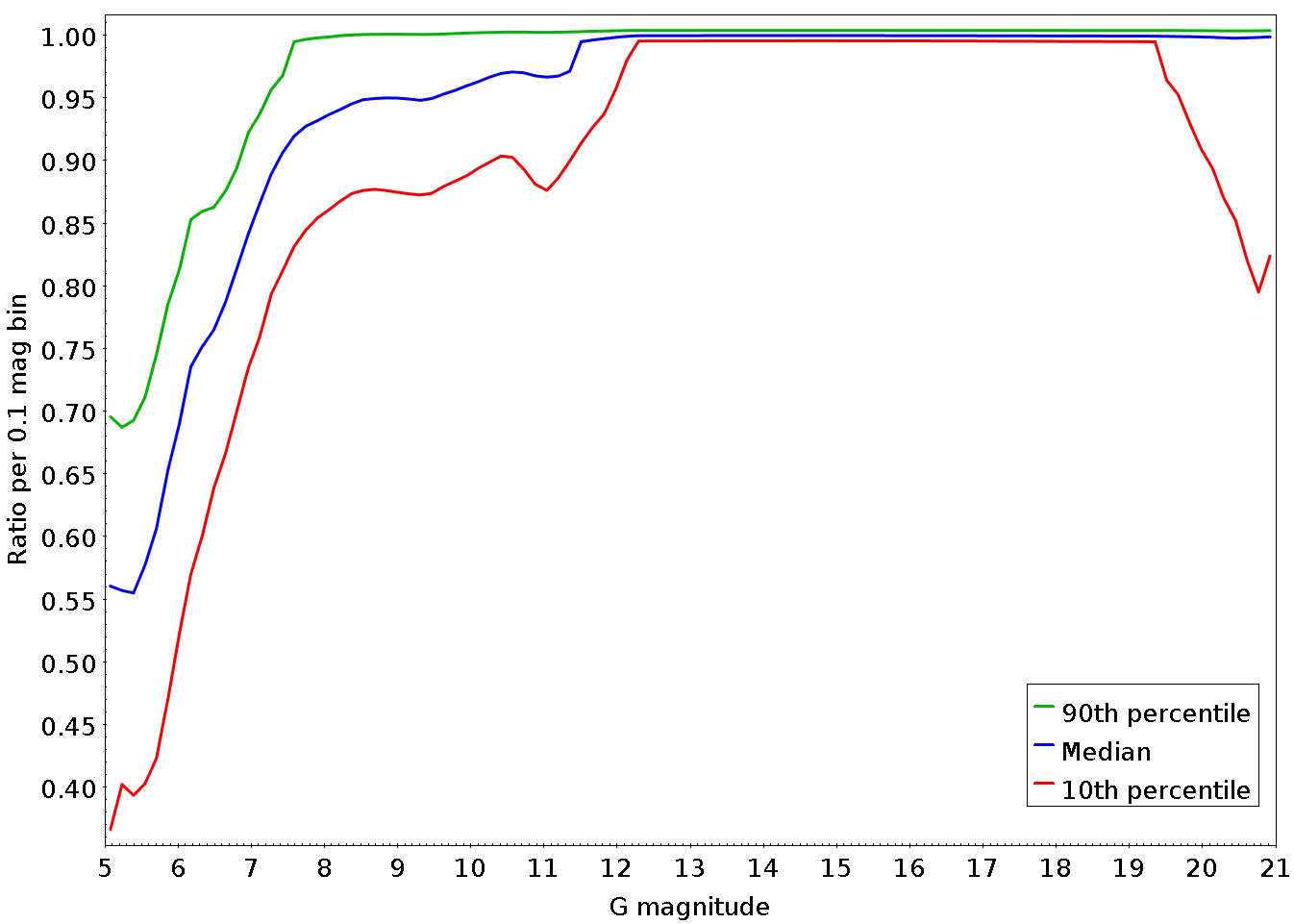}
  \caption{Fraction of matched transits that are retained from the \gdr2 XM solution by magnitude for the sources after the XM solution. The top horizontal green line shows the location of the 90th percentile, the blue line shows the location of the median, and the red line shows the location of the 10th percentile.}
  \label{fig:transitEvolution}
  \end{center}
\end{figure}

Figure~\ref{fig:transitEvolution} shows the fraction of matched transits that are retained from the \gdr2 XM solution by magnitude for the sources after the XM solution. There is a clear trend for both very bright and faint sources to lose more transits than for stars of medium magnitudes. Nevertheless, most of the bright transits are assigned to the same sources, and the total number of matched transits for the bright sources is incremented through the transits from the added period of time. 

\begin{figure}[h]
  \begin{center}
  \includegraphics[width=0.95\columnwidth]{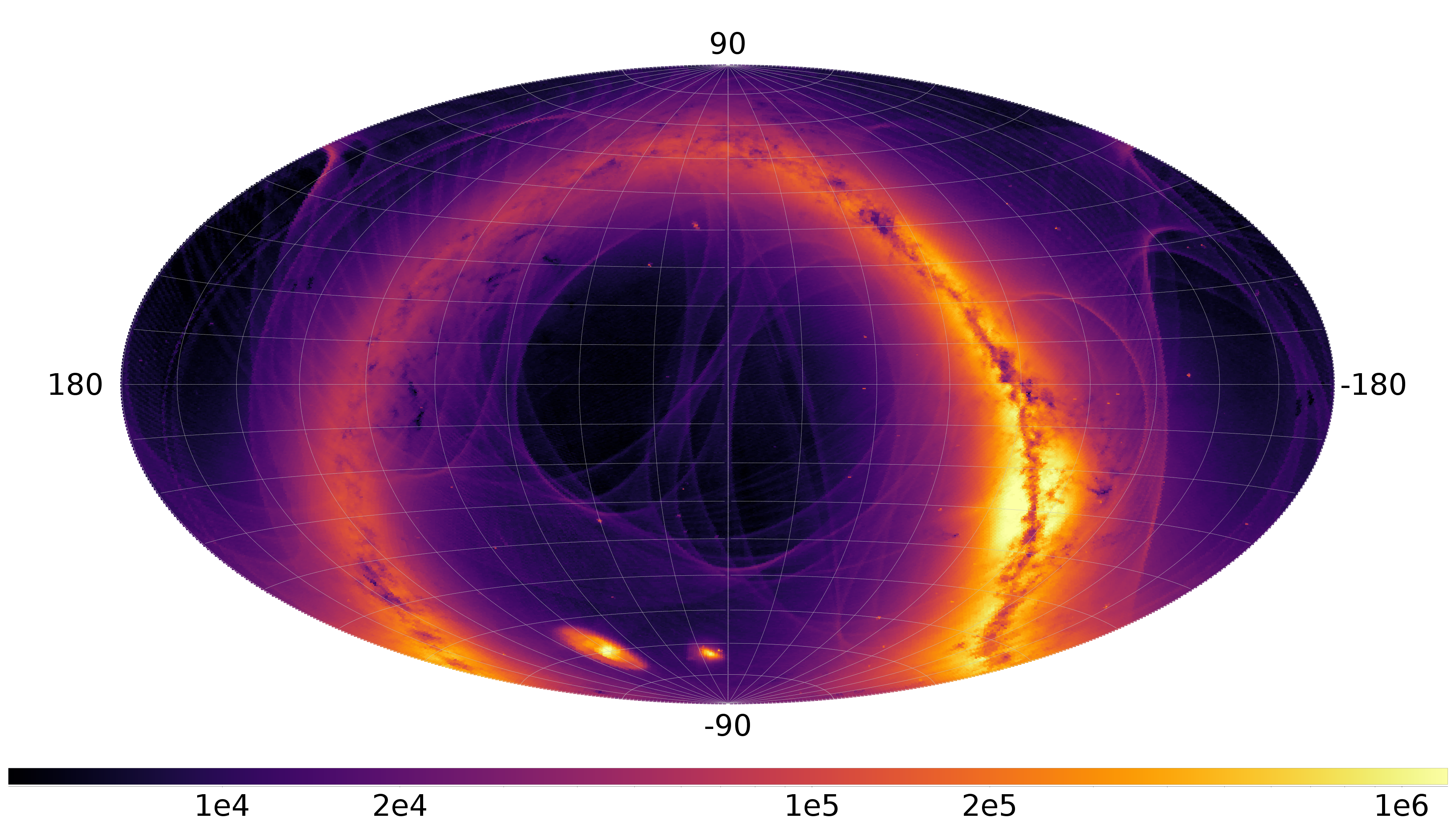}
  \caption{Density map of sources after the XM solution that retain more than 90\% of their matched transits from the previous XM at a pixel resolution of 0.21~deg$^2$.}
  \label{fig:MatchingTransitsRatioG90}
  \end{center}
\end{figure}

\begin{figure}[h]
  \begin{center}
  \includegraphics[width=0.95\columnwidth]{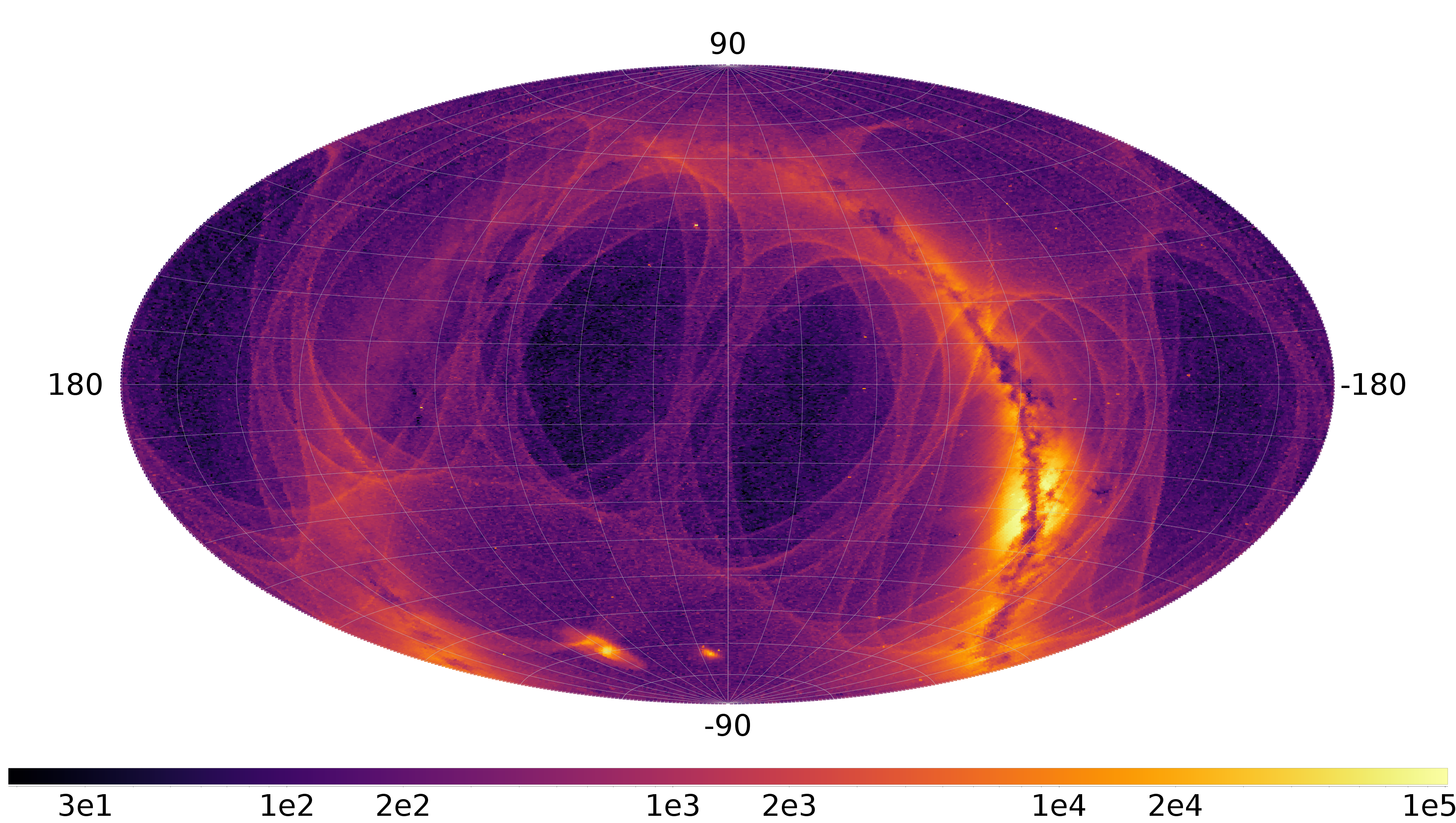}
  \caption{Density map of sources after the XM solution that retain less than 50\% of their matched transits from the previous XM at a pixel resolution of 0.21~deg$^2$.}
  \label{fig:MatchingTransitsRatioL50}
  \end{center}
\end{figure}

Figure~\ref{fig:MatchingTransitsRatioG90} shows that the distribution of sources after the XM solution that retain more than 90\% of their matched transits from the \gdr2 XM solution follows the natural distribution of sources on the sky with high star densities in regions such as the Galactic centre. On the other hand, Fig.~\ref{fig:MatchingTransitsRatioL50} suggests that the sources that loose the main part of their matched transits are more likely to be in areas with a larger number of matched transits or spurious detections. We note that the location of the spurious sources and spurious detections depends on the \gaia\ scanning law. Their sky density distribution is shown in Fig. \ref{fig:dcMap}. It is reasonable to expect this behaviour because the spurious detections and the spurious sources of the input catalogue may increase the number of candidate cluster-source links and the complexity of the algorithm.

\subsection{High proper motion sources}

For \gdr2, a linear source model was used in the clustering algorithm for sources with proper motions greater than 1000~mas~yr$^{-1}$. This implied a significant improvement with regard to \gaia~DR1, but some HPM stars were still missing. For \egdr{3}, the clustering algorithm takes proper motions for all the groups of detections into account, as discussed in Sect.~\ref{S:clustering}. Therefore the results for HPM stars improve even more for \egdr{3} by the updates in the clustering model and the increase in the time span. We recall that the final treatment of the HPM sources is performed by the astrometric solution \citep{EDR3-DPACP-128} and depends on the filters of the five-parameter astrometric solution.
This section considers the statistics over the set of HPM sources with a proper motion larger than 600~mas~yr${^{-1}}$.

According to the source evolution, most of the new HPM sources found in the XM are created by merging (about 8\%), and to a lesser extent, from scratch (about 1\%). These statistics are similar for different magnitude ranges but are more significant for bright sources, as shown in Table~\ref{tb:hpmevolution}.

\begin{table}[t]
  \caption{Statistics of HPM sources (proper motion larger than 600~mas~yr${^{-1}}$) for \egdr{3} for different magnitude ranges.}
\small
\centering
  \begin{tabular}{crrr}
  \hline\hline
    \noalign{\smallskip}
  Magnitude & New merge & New from scratch & Persisting \\ 
    \noalign{\smallskip}
\hline
    \noalign{\smallskip}
$G < 13$ & 13.2\% & 2.3\% & 84.5\% \\ 
$G < 18$ & 8.5\% & 1.2\% & 90.3\% \\ 
  \noalign{\smallskip}
Full & 8.3\% & 1.1\% & 90.6\% \\ 
  \noalign{\smallskip}
\hline
  \end{tabular}
  \medskip
  \label{tb:hpmevolution}
\end{table}

The total number of HPM sources for \egdr{3} is 2729, which is about 26\% less than in \gdr2. The reason of this is the large number of spurious sources with a HPM value and $G>18$ in \gdr2. That is, about 41\% of these faint HPM sources had a significant negative parallax and fewer than eight visibility periods in \gdr2. Symmetrically, about 28\% of these faint HPM sources had a parallax larger than 100~mas and fewer than eight visibility periods. The post-processing algorithm (explained in Sect.~\ref{SS:post-processing}) ensured that no HPM sources have a negative parallax in \egdr{3}. Therefore the number of HPM sources has decreased by about 85\% at the faint end ($G>18$), whereas the total number of HPM sources has increased for the bright sources: about 2\% for the sources with $G<18$ and about 5\% for the sources with $G<13$.

\begin{figure}[h]
  \begin{center}
  \includegraphics[width=0.95\columnwidth]{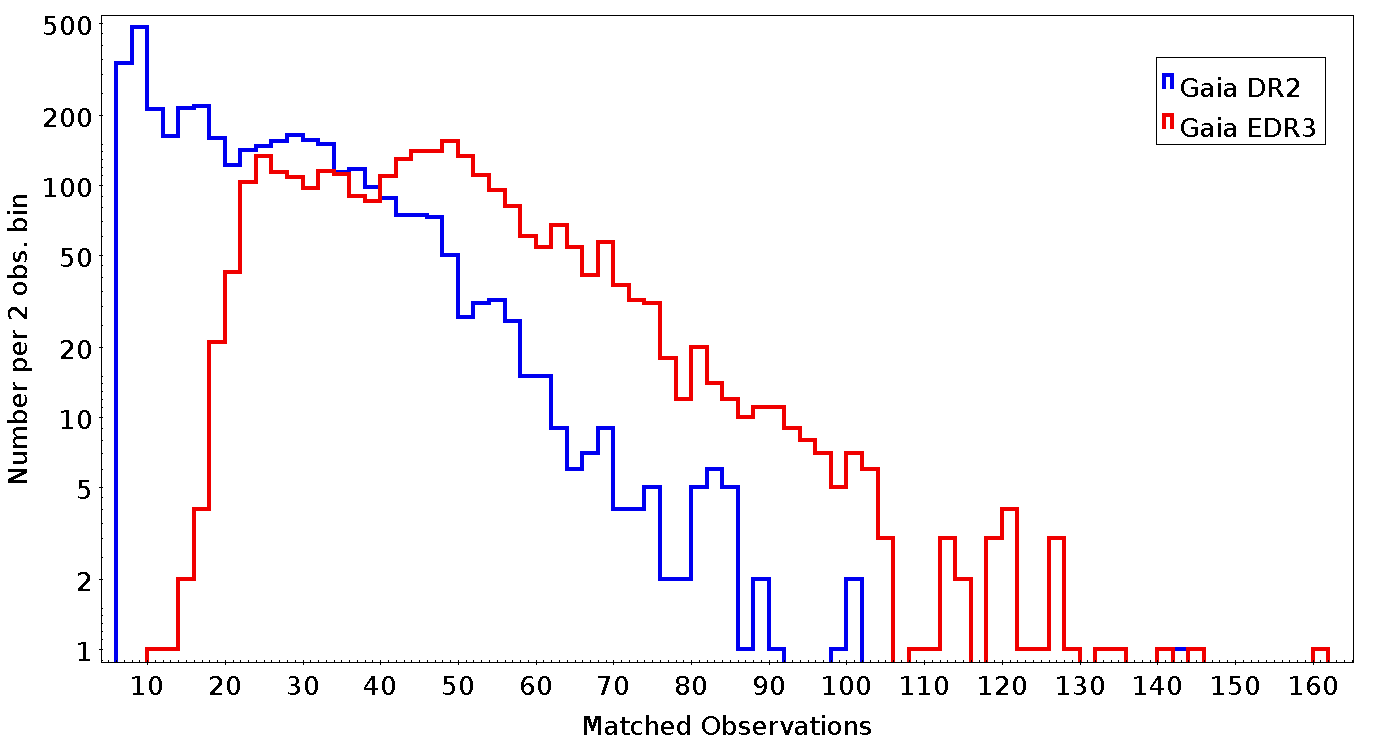}
  \caption{Distribution of matched transits for HPM sources (proper motion larger than 600~mas~yr${^{-1}}$). \egdr{3} contains data from 12 more months.}
  \label{fig:C03HPMSourceMatches}
  \end{center}
\end{figure}

The improved clustering model also affects the number of matched transits of HPM sources in \egdr{3}. Figure~\ref{fig:C03HPMSourceMatches} shows the number of matched transits of HPM sources. More than 50\% of the sources have at least 45 matched transits. Compared with Fig.~\ref{fig:C03IntegratedSourceMatches}, we can observe a similar behaviour of the number of matched transits. Thus, this parameter does not depend on the value of the proper motion, in contrast to \gdr2. 

The comparison with the number of matched transits of the data segments already used in \gdr2 is more interesting. About 17\% of the persisting HPM sources increase the number of matched transits from past segments while retaining all matched transits from \gdr2. In this way, the total number of HPM sources with at least 15 matched transits from past segments has increased by about 1\% with regard to \gdr2.

In conclusion, the HPM sources of \egdr{3} contain more matched transits and more accurate parameters because of the improved clustering model of the XM. Moreover, the post-processing analysis (Sect.~\ref{SS:post-processing}) reduced the spurious HPM sources. The completeness of the published HPM sources also depends on the posterior treatment given by the astrometric pipeline and the filters in the subsequent processes, described in \cite{EDR3-DPACP-128}.
%
%-------------------------------------------------------------------

\section{Conclusions and future improvements}\label{S:conclusions}
%Detail pending issues, caveats, limitations... such as close pairs and 
%duplicated sources from spurious detections.

We have described the XM solution implemented for the third \gaia\ data release with 78 billion processed transits. This solution is crucial for the performance of both astrometric and photometric parameters because it provides the matched observations that are to be used in the subsequent pipelines within the \textit{Gaia} DPAC. 

Compared to \gdr2, the detection classification was improved significantly around bright sources with the revision of the transit classification model parameters. Moreover, an additional module has been included that provides a cluster classification based on the quality of the transit windows. It prevents the creation of new sources when most of the cluster transits are dubious.

As detailed in this paper, a novel generalisation of the clustering model based on the nearest-neighbour chain algorithm has been developed. This new implementation improves the results for HPM sources and variable stars and also provides a much cleaner catalogue. Based on the larger number of observations and the post-processing analysis, we have obtained a reduction in the number of sources with false but significant negative or large parallaxes and in the number of source pairs that are separated by less than 400~mas as compared to \gdr2. These sources were considered duplicated sources for \gdr2, and this separation limit is reduced for \egdr{3}  to 180~mas through the cleaned catalogue that is provided by the XM and the astrometric solution. The reduction of spurious sources is higher in dense areas such as the Galactic centre, but some spurious sources still remain.

In this release, the source evolution is stabilised with far fewer changes in the source list between \gdr2 and \egdr{3} than between \gdr1 and \gdr2; 97.5\% of the sources persist. The updated merging and split criteria reduce the number of superseded as well as new sources. However, the improvements in the previous stages (described in Sect.~\ref{S:clustering}) may cause an evolution of the catalogue for specific cases. This is the case for the HPM sources and variable stars because of the developments in the clustering algorithm, or for the bright sources around magnitude $G = 10$ because the double-detection threshold was recalibrated. Moreover, an evolution may also occur for very bright sources where the large number of spurious detections and spurious sources may create multiple cluster-source links that can lead to new source identifiers for these bright sources. Other sources may also  evolve for many different reasons (e.g. newly rejected detections or more observations added in this release), but they represent a small fraction of the full catalogue.

Some source identifiers are not maintained across different cycles, and the astrometric and photometric parameters of a surviving source may be updated between different releases. Our recommendation therefore is to treat the source identifiers in each release as being from completely independent catalogues. A table tracing the sources from \gdr2 to \egdr{3}, \dt{gaiaedr3.dr2\_neighbourhood}, is provided in \gaia\ archive in order to easily identify \gdr2 sources in \egdr3.

For future data releases, the inclusion of the parallax as well as other types of source parameters may be taken into account in the clustering model. When the parallax is included in the source model, the current stopping rule for the agglomerative algorithm may be updated and may no longer require any threshold dependence on the parallax.

In spite of the improvements of the spurious parallaxes and duplicated sources, some remain in \egdr{3}. The precision of the input data in the XM is not good enough to detect and resolve close source pairs with a large number of matched detections. We therefore expect that the vast majority of the significant negative parallaxes found by the astrometric pipeline \citep{EDR3-DPACP-128} are in fact caused by disturbances from surrounding sources. A similar population of positive parallaxes will also be distorted, but be more difficult to identify. We also speculate that the majority of cases in which the astrometric solution falls back to
two-parameter solutions for sufficiently bright sources with plenty of good observations and a sufficient number of visibility periods are caused by disturbances from neighbours.

For future releases the XM could therefore consider that the onboard detections may include more than one peak in the detected CCD image. In this way, the XM could include information from the IPD and thus assign sources not only to clusters of onboard detections, but also to the individual peaks when sources are resolved sufficiently often in the IPD. Thus, the XM will derive celestial coordinates from the peaks in order to identify which peaks belong to which source after the clustering stage. For secondary peaks in 1D windows, there will be no AC information, and we will need a non-trivial range of scan angles to derive a reliable position. 

In summary, the XM of observations in \egdr{3} is more stable and reduces the number of spurious sources from 20.7\% to 16.1\%. It also solves other known issues of the previous releases that were due to suboptimal XM solutions. The main improvements for future releases in the XM solution will be related to the systematic IPD search of multiple peaks in each window, which will allow distinguishing close source pairs in the XM solution.

%--------------------------------------------------------------------

\begin{acknowledgements}
This work has made use of data from the ESA space mission Gaia, processed by the Gaia
Data Processing and Analysis Consortium (DPAC).
Funding for the DPAC has been provided by national institutions, in particular the institutions
participating in the Gaia Multilateral Agreement. The Gaia mission website is: \url{http://www.cosmos.esa.int/gaia}.

The authors are members of the DPAC, and this work has been supported by the following funding agencies: 

the Spanish Ministry of Economy (MINECO/FEDER, UE) through grants ESP2016-80079-C2-1-R, RTI2018-095076-B-C21 and the Institute of Cosmos Sciences University of Barcelona (ICCUB, Unidad de Excelencia `Mar\'{\i}a de Maeztu’) through grant MDM-2014-0369 and CEX2019-000918-M; 

the European Space Agency in the framework of the Gaia project and through ESA/ESTEC Contract No. 4000114776/15/NL/IB with DAPCOM Data Services S.L.;

the German Aerospace Agency DLR under grants 50QG0501 and 50QG1401;

The authors thankfully acknowledge the computer resources from MareNostrum, and the technical expertise and assistance provided
by the Red Espa\~nola de Supercomputaci\'on at the Barcelona Supercomputing Center, Centro Nacional de Supercomputaci\'on.
\end{acknowledgements}

\bibliographystyle{aa} % style aa.bst
\bibliography{refs} % your references refs.bib

\vfill
\begin{appendix}
\section{List of acronyms}
Below, we give a list of acronyms.\hfill\\
\begin{tabular}{ll}
\hline\hline 
    \noalign{\smallskip}
Acronym & Description \\
    \noalign{\smallskip}
\hline
    \noalign{\smallskip}
AC&Across-scan (direction) \\
AF&Astrometric field (CCDs) \\
AL&Along-scan (direction) \\
BCRS&barycentric celestial reference system \\
BP&blue photometer \\
CCD&charge-coupled device \\
DPAC&Data Processing and Analysis Consortium \\
DR1&\gaia\ Data Release 1\\
DR2&\gaia\ Data Release 2\\
DR3&\gaia\ Data Release 3\\
EDR3&\gaia\ Early Data Release 3\\
FoV&field of view \\
HEALPix&hierarchical equal-area iso-latitude pixelisation \\
HPM&high proper motion \\
ICRS&international celestial reference system \\
IGSL&initial \gaia\ source list \\
IPD&image parameter determination \\
LHS&revised Luyten half-second catalogue \\
RP&red photometer \\
RVS&radial velocity spectrometer \\
SM&sky mapper (CCDs) \\
XM&cross-matching \\
    \noalign{\smallskip}
\hline
\end{tabular} 
\end{appendix}
\end{document}